\documentclass[aps,prfluids,onecolumn,showpacs,showkeys,
longbibliography,superscriptaddress,amsmath,amssymb,floatfix,nofootinbib,a4paper]{revtex4-2}

\usepackage{mathtools}
\usepackage{graphicx}
\usepackage[usenames,dvipsnames]{color}
\usepackage{xcolor}
\usepackage{epsfig}
\usepackage{wasysym}
\usepackage{natbib}
\usepackage{bm}
\usepackage[normalem]{ulem}

\newcommand{\br}{{\bf r}}

\newcommand{\bj}{{\bf j}}
\newcommand{\be}{{\bf e}}
\newcommand{\bff}{{\bf f}}

\newcommand{\bv}{{\bf v}}

\newcommand{\bV}{{\bf V}}
\newcommand{\bu}{{\bf u}}

\newcommand{\bG}{{\bf G}}

\newcommand{\bA}{{\bf A}}
\newcommand{\bS}{{\bf S}}

\newcommand{\bOmega}{{\boldsymbol{\Omega}}}
\newcommand{\bomega}{{\boldsymbol{\omega}}}
\newcommand{\mulm}{\nu}

\newcommand{\eq}[1]{Eq.~(\ref{#1})}
\newcommand{\eqs}[1]{Eqs.~(\ref{#1})}

\newcommand{\rcite}[1]{Ref.~\cite{#1}}
\newcommand{\eref}[1]{(\ref{#1})}

\newcommand{\Fref}[1]{Figure~\ref{#1}}

\begin{document}

\title{Self-chemophoresis in the thin diffuse interface approximation}

\author{Alvaro Dom\'\i nguez}
\email{\texttt{dominguez@us.es}}
\affiliation{F\'\i sica Te\'orica, Universidad de Sevilla, Apdo.~1065, 
41080 Sevilla, Spain}
\affiliation{Instituto Carlos I de F{\'i}sica Te{\'o}rica y
  Computacional, 18071 Granada, Spain}

\author{Mihail N. Popescu}
\email{\texttt{mpopescu@us.es}}
\affiliation{F\'\i sica Te\'orica, Universidad de Sevilla, Apdo.~1065, 
41080 Sevilla, Spain}

\begin{abstract}
  Self-chemophoresis is an appealing and quite successful interpretation of the motility
  exhibited by certain chemically active colloidal particles suspended in a
  solution of their ``fuel'': the particle has a phoretic response to self-generated, rather than
  externally imposed, inhomogeneities in the chemical composition of
  the solution. The postulated mechanism of chemophoresis is the
  interaction of the particle (via an adsorption potential) with the
  chemical inhomogeneities in the surrounding medium. When the range
  of this interaction is much smaller than any other relevant scale in
  the system, the (translational and rotational) phoretic velocities
  can be described in terms of a phoretic coefficient and a slip fluid
  velocity at the surface of the particle.  Using the case of a spherical particle as a simple and physically insightful example, here we exploit an
  integral representation of the rigid-body motion to critically
  re-examine this framework. The  systematic analysis highlights two steps in the approximation: first
  the  thin--layer approximation (very large particle size), and
  subsequently a lubrication approximation (slow variation of the
  adsorption potential along the tangential direction). 
  We also discuss how these approximations could be relaxed and
  the effect of this on the particle's motion.
\end{abstract}


\maketitle

\section{Introduction}\label{sec:intro}

An important class of active matter systems consists 
of chemically active colloids. These particles can achieve
self-propulsion via the catalytic promotion on their surface of
chemical reactions in the surrounding solution (insightful reviews of
the many experimental realizations of such particles can be found in
Refs.~\cite{Ebbens2010,BDLR16,DEY2016,GWSS20}). Such systems serve as
benchmark examples for complex non-equilibrium steady-states
processes, e.g., the formation of ``living crystals''
\cite{Palacci2013}, or the self-assembly of rotating gears
\cite{Palacci2018,DiLeonardo2016}. From an application viewpoint, they
are envisioned, e.g., to act as ``carriers'' in portable lab-on-a-chip
devices \cite{Sen2008,Baraban2012,Wang2015} or to facilitate novel
methods of separation and purification \cite{Katuri2018,Bekir2023}.

The emergence of self-motility for chemically active colloidal
particles suspended in an aqueous solution of their ``fuel'' (e.g.,
hydrogen peroxide for a platinum-capped polystyrene particle) has been
the topic of many theoretical investigations since the first
experimental reports were published about two decades ago
\cite{Paxton2004,Fournier-Bidoz2005}.  It has been proposed that
self-phoresis is the mechanism for motility in many of the
experimental realizations, and not only in the case of chemically
active particles \cite{GLA05,Palacci2013,Simmchen2016}, but also as
self-chemophoresis via demixing a critical binary liquid mixture
\cite{Volpe2011} or as self-thermophoresis through a single-component
fluid \cite{Kroy2016}. Phoresis is a classic, very interesting example of non-equilibrium
thermodynamics and transport at zero Reynolds number: for a colloidal
suspension, an imposed thermodynamic force --- e.g., a gradient of
chemical potential or temperature in the solution --- may induce
motion of the particles and hydrodynamic flow of the solution due to
local force imbalances while in the absence of external forces or
torques, i.e., while the whole composed system ``particle + solution''
remains mechanically isolated \cite{DSZK47,Anderson1989}. In this
vein, self-phoresis is understood as a phoretic response to self-generated --- rather than externally
imposed --- gradients, so that the particles can be properly termed
``swimmers''. This interpretation, which was already envisioned by Anderson \cite{Anderson1989} four decades ago, is very appealing for its immediate connection with the vast body of
knowledge about phoresis, and it is now generally accepted as being
useful for understanding many of the features observed in experimental
realizations of such active particles
\cite{RuKr07,Golestanian2007,Moran2017,Kroy2016,Wurger2015,samin2015self,Brown2017,
  Koplik2013,Seifert2012,MiLa14,Popescu2016}. However, recent
developments \cite{DPRD20,Corato2020,DoPo22,Cruz2024,DoPoion24} are
challenging the paradigm that self-phoresis is phoresis in self-generated gradients, and
Refs.~\cite{DPRD20,DoPo22} showed that it breaks down for
self-chemophoresis in the simplest extension of the theory to account
for correlations in the solution (which turns out to also provide a
more general framework \cite{DoPoion24} for the reports in
Refs.~\cite{Corato2020,Cruz2024}). These conceptual developments,
together with the similarly interesting ones in phoresis related to,
e.g., phoresis in multivalent electrolytes
\cite{Stone2019b,Stone2020} or transversal salt gradients
\cite{Warren2020,Warren2024}, bring a novel impetus to the research in
this area.

In many cases, the interactions between the particle and the
ambient solution, which are lately responsible for ``converting'' the thermodynamic
gradients into fluid flow and particle motion, have a range much smaller than the size of the particle, so that they are
effectively restricted to a thin layer close to the particle's surface
(a ``diffuse interface'' in the language of
Ref.~\cite{Anderson1989}). By using chemophoresis (motion in gradients
of uncharged solutes, often also called ``diffusiophoresis'') as the
simplest conceptual example, it was proposed by Ref.~\cite{DSZK47}
that in such cases the dynamics in this thin layer can be modeled as a
local ``phoretic (often also called osmotic) slip'' velocity $\bv_s$
\textit{tangential to the surface of the particle} that plays the role
of a hydrodynamic boundary condition for the flow field $\bu(\br)$ of
the solution. This proposal was extended to a wide range of phoretic
phenomena, as discussed by the review of Anderson \cite{Anderson1989},
and then conjectured to apply to the case of self-phoresis in
Ref.~\cite{GLA05}. This latter conjecture was set on solid mathematical grounds for
self-chemophoresis in Ref.~\cite{MiLa14} by using a lubrication
approximation for the flow within the diffuse interface.

When the phoretic--slip description holds, the calculation of the translational and rotational velocities,
$\bV$ and $\bOmega$ respectively, of the rigid body motion of the
particle is greatly simplified. The use of the reciprocal theorems for
incompressible Stokes flows
\cite{BrennerBook,KiKa91,Lorentz_original,Lorentz_transl} provides
$\bV$ and $\bOmega$ as integrals over the surface of the particle of the phoretic slip
$\bv_s$ weighted by certain geometrical factors that are dependent
only on the shape of the particle \cite{Anderson1989}. In particular,
for a spherical particle of radius $R$ this approach renders the
following well known expressions: 
\begin{subequations}
  \label{eq:V_Om_aver_slip_sphere}
  \begin{equation}
    \label{eq:slips}
    \bV = - \langle \bv_s \rangle\,,
    \qquad
    \bOmega = - \left \langle \bomega_s \right\rangle ,
  \end{equation}
  where
  \begin{equation}
    \label{eq:oms_vs}
     \bomega_s = \frac{3}{2R} \be_r \times \bv_s ,
  \end{equation}
  and 
  \begin{equation}
    \label{eq:Saverage}
    \langle (\cdots) \rangle := \frac{1}{4 \pi R^2}
    \oint\limits_{|\br| = R} dS \,(\cdots) \,.
  \end{equation}
\end{subequations}

By using, for reasons of conceptual and technical simplicity, the case
of a self-chemophoretic spherical particle, we employ a recently derived integral representation for the rigid--body motion of the particle \cite{DPRD20,DP24,DoPo22} to re-examine the diffuse interface picture. This allows us to decouple two steps in the approximation: one starts with the thin--layer approximation (a thin diffuse interface), and subsequently follows
with the lubrication approximation proper (slow variations along the tangential direction). We analyze the conditions for realistic particles under which the thin layer approximation holds but the lubrication approximation does not apply and show that this may reflect quantitatively, but not qualitatively, in the resulting motion (at least not when the inhomogeneities represent a small departure from equilibrium).

\section{\label{sec:model} Model system: a self-chemophoretic spherical particle}

\begin{figure}[!t]
\begin{center}
  \includegraphics[width=0.9\textwidth]{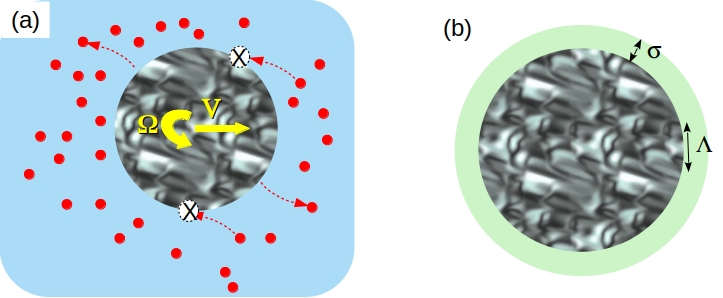}
\end{center}
\caption{ (a) Schematic depiction of the model system. A rigid,
  impermeable, spherical particle (the central disk), is immersed in a
  solution composed of solvent (the majority phase, shown as the blue
  background) and a single species of a molecular solute (the red
  small disks), which diffuses in the
  solution. 
  The particle is chemically active, i.e., it releases (or removes)
  solute molecules into (from) the solution adjacent to the
  surface. The texture of the particle is an illustrative depiction
  that its properties (adsorption potential, chemical activity)
  can vary along the surface. As a result of the coupling of the
  particle, via the adsorption potential, with the non-equilibrium
  inhomogeneities in chemical composition induced by the activity, the
  particle translates with phoretic velocity $\bV$ and rotates with
  phoretic angular
  velocity $\bOmega$ (yellow arrows), while the solution flows.
  (b) The range $\sigma$ of the adsorption potential defines a
  ``diffuse'' particle--fluid interface, where the interaction is
  effective. The lateral variations of particle properties along
  its surface are characterized by a length scale $\Lambda$.}
\label{fig:fig_schematic}
\end{figure}

The presentation of the self-chemophoresis framework follows closely
the one in Refs.~\cite{DPRD20,DoPo22}. We consider a rigid,
impermeable, spherical particle (radius $R$) suspended in a liquid
solution consisting of a solvent plus a single solute species (called
``the chemical'' in the following). The chemical diffuses in the
solution with diffusion constant $D$. The particle is chemically
active, i.e., it is a source or sink of the chemical (see the
schematic in \Fref{fig:fig_schematic}(a)). We assume constant
temperature $T$ and mass density of the solution, and --- in
accordance with the typical experimental observations, see, e.g.,
Ref.~\cite{Ebbens2010} --- a slow motion of the
particle. Consequently, the state of the solution is characterized by the
instantaneous stationary profiles of the number density $n(\br)$ of
the chemical at small P{\'e}clet numbers (i.e., flow advection of the solute is
neglected), and of the velocity field $\bu(\br)$ of the solution,
assumed to behave as a Newtonian fluid, at small Reynolds and Mach
numbers (i.e., creeping flow) \cite{deMa84}.

The field $n(\br)$ is determined by particle number conservation for the chemical,
\begin{equation}
  \label{eq:chemstat}
  \nabla \cdot\bj_\mathrm{diff} = 0 ,
  \qquad
  \bj_\mathrm{diff} = \Gamma \mathbf{f},
\end{equation}
in terms of the current $\bj_\mathrm{diff}$, the body force density
$\bff$ acting on the chemical and which drives its diffusion, and the
mobility $\Gamma = D/k_B T$ of the chemical in the
solution. Consistently with the slow dynamics of the
particle, one assumes local equilibrium for the chemical
\cite{Anderson1989,Wheeler1998,GLA05,Brown2017,samin2015self,Wurger2015},
so that
\begin{equation}
  \label{eq:f}
  \mathbf{f}(\br)= - n(\br) \nabla\mu(\br),
\end{equation}
in terms of its chemical potential $\mu(\br)$, modeled by means of a free energy functional \cite{Wheeler1998}:  
\begin{subequations}
  \label{eq:mu}
  \begin{equation}
    \mathcal{H}[n] = 
    \int\limits_{r>R} d^3\br \;
    \left[ 
      h(n) + n \, \mathbb{W}(\br) 
    \right] ,
  \end{equation}
  \begin{equation}
    \label{eq:mun}
    \mu(\br) = \frac{\delta\mathcal{H}[n]}{\delta n(\br)} 
    = h'(n(\br)) 
    + \mathbb{W}(\br) ,
  \end{equation}
\end{subequations}
where $h(n)$ is a local free energy density, which depends implicitly
also on temperature, and $\mathbb{W}(\br)$ is the potential energy
associated to the particle--chemical interaction. This potential has a non-zero and finite range and reflects the preference of the
particle for one of the components of the solution; it will be accordingly referred to as \textit{adsorption potential}.\footnote{For an incompressible solution, 
$\mathbb{W}$ is the difference between the interaction with the
particle experienced by a solute molecule at $\br$ and the one
experienced by a solvent molecule at the same point \cite{deMa84}.} We
note that the simplest choice for the free energy density $h(n)$,
frequently employed in models of phoresis and self-phoresis
\cite{Anderson1989,Golestanian2007,MiLa14}, is that of an ideal gas, 
\begin{equation}
  \label{eq:hideal}
  h_\mathrm{id}(n)=k_B T \,
  n(\ln n-1)\,.
\end{equation}

The diffusion equation that follows from \eqs{eq:chemstat} and
\eref{eq:f} is subject to boundary conditions at infinity and at the
surface of the particle. We impose that the perturbations due to the particle (through adsorption and
chemical activity) remain localized, so that the solution far from the
particle is at equilibrium, characterized by a homogeneous chemical
density $n_0$. At the surface of the particle, the chemical activity
of release/remove of chemical into/from the solution is modelled as a
chemical current 
along the direction normal to the particle. Accordingly,
\begin{subequations}
  \label{eq:BCs_diff}
\begin{equation}
  \mu(|\br| \to \infty) \to \mu_0 = h'(n_0)\,,
\end{equation}
\begin{equation}
  \be_r \cdot \bj_\mathrm{diff} (\br_p)
  =\mathcal{A}\, \mathbb{A}(\br_p)\,,
\end{equation}
\end{subequations}
where $\br_p$ generically denotes any point on the surface of the
particle. The parameter $\mathcal{A}$ is a measure of the magnitude of the chemical activity
(e.g., the maximum rate of chemical release or removal over the
surface), and the dimensionless surface field $\mathbb{A}(\br_p)$
is the pattern of surface chemical activity.

The flow field $\bu(\br)$ is assumed to be incompressible and determined by the balance between the force field $\bff(\br)$, which acts on the solution, and the fluid stresses, as expressed by the incompressible Stokes equation:
\begin{subequations}
  \label{eq:stokes}
  \begin{equation}
    \label{eq:localFbalance}
    \eta\nabla^2\bu - \nabla p
    + \mathbf{f} = 0 ,
  \end{equation}
  \begin{equation}
    \label{eq:incomp}
    \nabla\cdot\bu=0 ,
  \end{equation}
\end{subequations}
where $\eta$ is the viscosity of the solution and $p(\br)$ is the
hydrodynamic pressure field that enforces incompressibility. The equations are subject to boundary conditions: a vanishing velocity
at infinity, that sets the reference frame consistent with the
equilibrium state of the solution, and the usual no-slip at the surface of the particle:
\begin{subequations}
  \label{eq:BCs_Stokes}
\begin{equation}
  \bu(|\br| \to \infty) \to 0\,,
\end{equation}
\begin{equation}
  \bu(\br_p) = \bV + \bOmega \times \br_p\,.
\end{equation}
\end{subequations}
The quantities $\bV$ and $\bOmega$ are still unknowns. The
mathematical problem is closed by the requirement that the composed
system ``particle + solution'' is mechanically isolated, that is,
there is no external force or torque acting on the particle (recall that, as follows from \eq{eq:localFbalance} and the
  physical meaning of $\bff$ given by Eqs.~(\ref{eq:f}, \ref{eq:mu}),
there are no external forces acting on the fluid). One can then solve
\eqs{eq:chemstat}--\eref{eq:BCs_Stokes} in order to derive the
velocities $\bV$ and $\bOmega$ in terms of the force field~$\bff$. 

However, one can sidestep the need to solve the full
hydrodynamic problem~(\ref{eq:stokes}--\ref{eq:BCs_Stokes}) for the
flow $\bu$ by applying the reciprocal theorem
\cite{Teubner1982,Seifert2012,DPRD20,DP24}, which provides $\bV$ and
$\bOmega$ directly as functionals of the force
density~(\ref{eq:f}). The incompressibility
constraint~(\ref{eq:incomp}) implies that only the solenoidal
component of the force density matters\footnote{From
  \eq{eq:localFbalance}, it follows that any potential component
  $\nabla \chi$ of the force $\mathbf{f}(\br)$ is absorbed by the
pressure field $p(\br)$.}. Therefore, instead of the customary expression
for $\bV, \bOmega$ in terms of the field $\mathbf{f}(\br)$, we follow
the approach recently proposed in \rcite{DP24} of using an expression
that depends explicitly just on $\nabla\times\mathbf{f}(\br)$: this
has the advantage that any approximation in the integral representation (which will be unavoidable for further analytical progress or in numerical implementations with finite precision) does not inadvertently carry over a spurious contribution to the motion from a non-solenoidal component of the force. One then introduces one scalar and three vector hydrodynamic potentials, which for a
spherical particle take the following form, respectively:
\begin{subequations}
  \label{eq:QAspher}
  \begin{equation}
    \label{eq:Qspher}
    \Phi(r) = - \frac{3}{2} R^2 \left[ 1 - \frac{2}{3} \frac{R}{r} -
      \frac{1}{3} \left(\frac{r}{R}\right)^2 \right] ,
  \end{equation}
  \begin{equation}
    \bA^{(k)} (\br) = A(r) \, \be_k\times\be_r ,
  \end{equation}
  \begin{equation}
    \label{eq:Aspher}
    A(r) := 
    \frac{3}{4} R \left[ 1 - \frac{2}{3} \frac{r}{R} -
      \frac{1}{3} \left(\frac{R}{r}\right)^2 
    \right] , 
  \end{equation}
\end{subequations}
in spherical coordinates with the origin at the center of the particle
($\be_k$, $k=1,2,3$, are Cartesian unit vectors, and $\be_r$ is the
unit radial vector), for which we chose the gauge such that the potentials vanish at the surface of the particle and the fields
$\bA^{(k)}$ are solenoidal \cite{DP24}. By using this approach, one
obtains a representation like in \eq{eq:slips} with the following
exact integral representations of the surface fields
\cite{DPRD20,DP24}:
\begin{subequations}
  \label{eq:V_Om_curlf}
  \begin{equation}
    \label{eq:vs_exact}
    \bv_s(\br_p) = - \frac{2}{3\eta R} \int\limits_{0}^{\infty} dz\; (R+z)^2\,
    A(R+z) \,\be_r \times \left[ \nabla \times \bff(\br_p+z\be_r) \right],
  \end{equation}
  \begin{equation}
    \label{eq:oms_exact}
    \bomega_s(\br_p) = - \frac{2}{\eta R^3}
    \int\limits_{0}^{\infty} dz\; (R+z)^2\,
    \Phi(R+z) \, \nabla \times \bff(\br_p+z\be_r) ,
  \end{equation}
\end{subequations}
where $z>0$ is the distance from the point $\br_p=R\be_r$ at the
surface of the particle (i.e., any point $\br$ in the fluid is
expressed as $\br = (R+z) \be_r$). Notice that the expression~(\ref{eq:vs_exact}) gives a velocity $\bv_s$ that is obviously tangential to the surface (the vector $\be_r$ does not depend on $z$). The direction of the other surface field, $\bomega_s$, is not evident and one cannot \textit{a priori} rule out a component along the normal to the surface; however, it turns out that its contribution to the surface
average~(\ref{eq:V_Om_aver_slip_sphere}) that gives $\bOmega$ involves solely tangential components, see, c.f.,
Eqs.~(\ref{eq:Om_lm}, \ref{eq:Gtau}). What one can nevertheless show
rigorously\footnote{For instance, it suffices to consider the case
  $\bff(\br)=f(r) \be_\varphi(\theta,\varphi)$ with a function $f(r)$
  such that $f(0)=0$ and $f(r\to\infty)\to 0$ sufficiently fast.} is
that these exact expressions for the fields $\bv_s$ and $\bomega_s$
are not related like in \eq{eq:oms_vs}.

The expressions~(\ref{eq:V_Om_curlf}) are generally valid irrespective of the exact form of the force field $\bff$ or the free energy functional $\mathcal{H}$, see Eqs.~(\ref{eq:f}, \ref{eq:mu}), i.e., they hold equally well for more complex mechanisms such as the correlation--induced phoresis \cite{DPRD20}. For the specific form given by \eq{eq:mu}, which depends only locally on $n$ via the free energy density $h(n)$, one has\footnote{We assume that the state of the system is sufficiently far from any phase transition, so that $h''(n)\neq 0$ for the values of $n$ that are of interest in the current problem.} 
\begin{equation}
  \label{eq:mu2}
  \nabla n = \frac{1}{h''(n)} \nabla\left(\mu-\mathbb{W}\right),    
\end{equation}
so that it holds 
\begin{equation}
\label{eq:curl_f_id}
  \nabla \times \bff
  = -\nabla n \times \nabla \mu
  = \frac{1}{h''(n)} \nabla \mathbb{W} \times \nabla \mu \, .
\end{equation}
This expression explicitly highlights that phoretic motion arises by
the misalignment between the adsorption force ($\sim\nabla\mathbb{W}$)
exerted by the particle on the solute molecules and the
non-equilibrium thermodynamic force ($\sim\nabla\mu$) due to the
chemical activity.

\section{\label{sec:thin_film} The thin--layer and the lubrication approximations}

As discussed earlier, in many cases the force density $\bff$ is
non-vanishing only in a thin layer at the surface of the particle (the
``diffuse interface'' \cite{Anderson1989}). For the form given by
\eq{eq:curl_f_id}, this usually happens when the adsorption potential
$\mathbb{W}$ has a range $\sigma$ that is much smaller than the
particle size, $\sigma\ll R$. In such case, the radial integrals in
\eqs{eq:V_Om_curlf} into the fluid domain are effectively cut off
($z=r-R\lesssim \sigma$) to the layer of fluid within the spherical
shell of thickness $\sim \sigma$ around the surface of the particle,
see \Fref{fig:fig_schematic}(b). Then, one can Taylor--expand the hydrodynamic kernels
$\Phi(r), A(r)$ (and the term $(R+z)^2$ coming from the volume
element) in the variable $z\ll R$ because they only involve the large scale
$R$. In this manner, one arrives at the \emph{thin--layer approximation} for
the phoretic velocities:
\begin{subequations}
  \label{eq:V_Om_thin_layer}
  \begin{equation}
    \label{eq:vs_thin}
    \bv_s(\br_p) \approx - \frac{2}{3} \br_p \times \bomega_s(\br_p) ,
  \end{equation}
  \begin{equation}
    \label{eq:oms_thin}
    \bomega_s(\br_p)
    \approx
    - \frac{3}{4 \eta R} 
      \int\limits_0^\infty dz\; z^2 \, \nabla \times \bff(\br_p + z
      \be_r) .
  \end{equation}
\end{subequations}
Notice that (i) although the integral over $z$ is formally extended to
infinity, the fast decay of $\bff$ ensures that only the thin--layer
domain contributes, and (ii) because the layer thickness is microscopic compared to the particle size, the fields $\bv_s(\br_p)$ and 
$\bomega_s(\br_p)$ can be properly called ``surface fields''.

So far, the approximation addresses only the behavior of the
hydrodynamic kernels based on the scale $\sigma$ of radial variation
of the field $\nabla\times\mathbf{f}$. In order to make further
progress, we have to study also its variation in the lateral
(tangential) direction on the particle's surface. Let $\Lambda_\mathbb{W}$ denote the characteristic scale of lateral
variation of the adsorption potential. Introduce the tangential
gradient $\nabla_\parallel := \nabla - \be_r \partial_r$, so that one can write
\begin{equation}
  \label{eq:gradW}
  \nabla\mathbb{W} = \nabla_\parallel \mathbb{W} + \be_r
  \frac{\partial \mathbb{W}}{\partial r} .
\end{equation}
One can thus estimate
\begin{equation}
  \frac{\left|\partial_r \mathbb{W}\right|}{\left|\nabla_\parallel
      \mathbb{W}\right|}
  \sim \frac{\Lambda_\mathbb{W}}{\sigma} .
\end{equation}
Therefore, an additional, natural approximation occurs when the
lateral variation is slow compared to the variation in the radial
direction, i.e., $\sigma\ll\Lambda_\mathbb{W}$, so that
\begin{equation}
  \label{eq:lubrW}
  \nabla \mathbb{W} \approx \be_r
 \frac{\partial \mathbb{W}}{\partial r} ,
\end{equation}
and \eq{eq:V_Om_thin_layer} are further simplified, upon using
\eq{eq:curl_f_id}, as
\begin{subequations}
  \label{eq:voms_lubr1}
\begin{equation}
  \label{eq:oms_lubr1}
    \bomega_s(\br_p)
    \approx
    - \frac{3}{4 \eta R} 
    \int\limits_0^\infty dz\; z^2 \, \frac{1}{h''(n)}
    \frac{\partial \mathbb{W}}{\partial z}
    \be_r \times \nabla_\parallel \mu ,
  \end{equation}
\begin{equation}
  \label{eq:vs_lubr1}
    \bv_s(\br_p)
    \approx
    - \frac{1}{2 \eta} 
    \int\limits_0^\infty dz\; z^2 \, \frac{1}{h''(n)}
    \frac{\partial \mathbb{W}}{\partial z}
    \nabla_\parallel \mu.
  \end{equation}
\end{subequations}
Now it is obvious that both fields are tangential and also fulfill
\eq{eq:oms_vs}, so that it suffices to focus on the tangential slip
velocity $\bv_s$ from now on.

At this point, a specific expression of the chemical potential is
needed. Unfortunately, even for the simplest case of an ideal
gas~(\ref{eq:hideal}), the diffusion problem (\ref{eq:chemstat}--\ref{eq:BCs_diff}) leads to a complicated partial differential equation that cannot be solved analytically. Accordingly, we now address the case of small deviations from the homogeneous, equilibrium state, for which the governing
equations can be simplified.

\subsection{\label{sec:quasi-hom} The quasi-homogeneous regime}

When both $\mathbb{A} = 0$ and $\mathbb{W} = 0$, the solution represents
an equilibrium state characterized by a uniform density $n_0$.
Therefore, if $\mathbb{A}$ and $\mathbb{W}$ are sufficiently small, one can linearize the governing equations by expanding to first order
in the deviations $\delta n:= n-n_0$. The diffusion problem reduces to a Laplace boundary value problem
for $\delta \mu:= \mu-\mu_0$:
\begin{subequations}
  \label{eq:bvp_dmu}
  \begin{equation}
    \label{eq:muLaplace}
    \nabla^2 \delta \mu = 0, 
  \end{equation}
  \begin{equation}
    \delta \mu (|\br| \to \infty) = 0, 
  \end{equation}
  \begin{equation}
    \label{eq:neumannMu}
    \be_r \cdot \nabla \delta \mu(\br_p)
    = - \frac{\mathcal{A}}{n_0 \Gamma} \mathbb{A}(\br_p) ,
  \end{equation} 
\end{subequations}
and linearization of \eq{eq:mu2}  then provides 
\begin{equation}
\label{eq:n_dev}
\delta n(\br) = \frac{1}{h''(n_0)} \left[ \delta \mu(\br) -
  \mathbb{W}(\br) \right] \,.
\end{equation}
Within this approximation, one can set
\begin{equation}
  \label{eq:h0}
  h''(n)\approx h''(n_0)   
\end{equation}
in the integrals appearing in \eqs{eq:voms_lubr1} and the phoretic
velocities appear, to leading order, as quadratic magnitudes in the
deviations from the equilibrium state.

For the spherical particle, \eqs{eq:bvp_dmu} can be solved
straightforwardly by expanding the fields in spherical harmonics: if the activity is expressed in terms of dimensionless coefficients $a_{\ell m}$ as
\begin{equation}
  \label{eq:alm}
  \mathbb{A}(\br_p=R \be_r) = \sum_{\ell m} a_{\ell m} Y_{\ell m} (\be_r) ,
\end{equation}
one gets
\begin{subequations}
  \label{eq:mulm}  
\begin{equation}
  \delta\mu(\br=(R+z)\be_r) =  \frac{\mathcal{A} R}{n_0 \Gamma}
  \sum_{\ell m}
  \mulm_{\ell m} (z)
  \, Y_{\ell m}(\be_r) .
\end{equation}
in terms of the dimensionless coefficients
  \begin{equation}
    \label{eq:mucoeff}
    \mulm_{\ell m}(z)
    := \frac{a_{\ell m}}{\ell+1} 
    \left(1+\frac{z}{R}\right)^{-(\ell+1)} .
  \end{equation}
\end{subequations}
The relevant feature is that the derivatives of the chemical potential
in the radial and in the tangential direction are of the same order
(this follows directly from the Laplace equation~(\ref{eq:muLaplace}),
but it can be checked also with the explicit
solution~(\ref{eq:mulm}), see App.~\ref{app:mu}). Therefore, if the
activity pattern $\mathbb{A}(\br_p)$ has a characteristic scale
$\Lambda_\mathbb{A}$ of lateral variation on the particle's surface,
so will $\delta\mu$ (via the common coefficients $a_{\ell m}$
appearing in Eqs.~(\ref{eq:alm}, \ref{eq:mulm})), and thus $\Lambda_\mathbb{A}$ will also set the scale of radial variation of
$\delta\mu$.

\subsection{\label{sec:lubr} The phoretic coefficient}

Assume now that, as happens with the length $\Lambda_\mathbb{W}$, also
the length scale $\Lambda_\mathbb{A}$ is much larger than the
thickness $\sigma$ of the diffuse interface. Then, in view of the
feature just remarked about the field $\delta\mu(\br)$, one can
approximate this field by its value at the surface of particle when
evaluating the integral in $z$ appearing in
\eqs{eq:voms_lubr1}. Therefore, together with the
approximation~(\ref{eq:h0}), one finally gets 
\begin{subequations}
  \begin{equation}
    \label{eq:vs_lubr2}
    \bv_s(\br_p) \approx \frac{\mathcal{L}(\br_p)}{h''(n_0)}
    \nabla_\parallel \delta\mu(\br_p) ,
  \end{equation}
\end{subequations}
  in terms of the coefficient
  \begin{eqnarray}
    \label{eq:phoreticL}
    \mathcal{L}(\br_p)
    & :=
    & \mbox{} - \frac{1}{2\eta}\int\limits_0^\infty dz\; z^2\,
      \frac{\partial \mathbb{W}}{\partial z}(\br_p+z\be_r)
      \nonumber
    \\
    & =
    & \frac{1}{\eta}\int\limits_0^\infty dz\; z\,
      \mathbb{W}(\br_p+z\be_r) ,
  \end{eqnarray}
  after integration by parts.  This result can be properly termed the \textit{lubrication
    approximation}, see Sec.~\ref{sec:discussion}.
  
  In order to illustrate the meaning of $\mathcal{L}$, we notice that
  it is experimentally much simpler to record gradients in density
  than in chemical potentials: when evaluating fields \textit{outside
    of the diffuse interface but still close to the particle}, i.e., at some
  value of $z$ such that $\sigma\ll z \ll R$, one can set
  $\mathbb{W}=0$ and, in particular, \eq{eq:n_dev} allows one to
  define the ``outer density'' as
  \begin{equation}
    \label{eq:nout}
    \delta n^\mathrm{(o)}(\br) := \frac{\delta\mu (\br)}{h''(n_0)} .
  \end{equation}
  Using this relationship to simplify Eqs.~(\ref{eq:vs_lubr2}), one
  finally arrives at
  \begin{equation}
    \label{eq:vs_lubr_quasihom}
    \bv_s(\br_p) = \mathcal{L}(\br_p) \, \nabla_\parallel \delta n^{\mathrm{(o)}}(\br_p)\,.
  \end{equation}
  One recognizes here the classic expression for the phoretic slip  velocity as proportional to the observable ``outer'' tangential
  gradient in chemical concentration (be it due to activity or
  externally imposed), where $\mathcal{L}$ is the phoretic coefficient
  \cite{DSZK47,Anderson1989,MiLa14}.
    Note that \eq{eq:vs_lubr_quasihom} corresponds to the classic result
  derived for an ideal gas \cite{DSZK47,Anderson1989},
  \begin{equation}
    \label{eq:Lclassic}
    \mathcal{L}_\mathrm{id} = \frac{k_B T}{\eta} \int\limits_{0}^{\infty} dz\; z \left[
      1 - \exp(-\mathbb{W}/k_B T)
      \right] ,
  \end{equation}
  in the quasi--homogeneous regime limit $|\mathbb{W}| \ll k_B
  T$; we remark in passing that this result can be derived from
    \eqs{eq:voms_lubr1} before implementing the approximation of
    quasi-homogeneity if one uses the insight that $\mu$ varies little
    in the radial direction when the conditions for the lubrication
    approximation hold, see App.~\ref{app:idealgas}.

  \section{\label{sec:discussion} Beyond the approximations}

The lubrication approximation is distinctly different from the
thin--layer approximation~(\ref{eq:V_Om_thin_layer}). In the latter,
only the curvature of the particle surface is neglected, so that the
hydrodynamic flow within the diffuse interface is approximated as if
near a flat wall. In the lubrication approximation also the lateral
variations of this flow along the surface are neglected, so that
the problem is reduced to a shear flow within the diffuse
  interface and parallel to the particle.

More precisely, the lubrication approximation amounts to claiming that
the thickness of the diffuse interface, set by the range $\sigma$ of
the adsorption potential in our case, is much smaller than any other
length scale that might be relevant for the hydrodynamic flow, be it
the curvature radius of the particle or the scale of lateral variation
of the fields. It seems natural to explore the possibility of relaxing
some of these constraints, particularly in view of the ever improving
technical capabilities in the fabrication of active particles. For
instance, it is currently feasible to get particles of radii in the
range $R \simeq 100~\mathrm{nm}$ \cite{Fischer2017,Sanchez2021}, for
which a pattern of alternating active/inactive surface over an angle
$\pi/16 (\sim 11^\mathrm{o})$ renders
$\Lambda_\mathbb{A}, \Lambda_\mathbb{W} \sim R/5 \simeq
20~\mathrm{nm}$; with a typical value $\sigma \simeq 5~\mathrm{nm}$,
the ratio $\Lambda_\mathbb{A,W}/\sigma \sim 4$ is quantitatively
significant and it cannot be taken for granted as
``very large'',  so that it could eventually lead
to observable discrepancies from the prediction based on
\eq{eq:vs_lubr_quasihom}.

In order to address the case when the simplifying approximations
discussed above are relaxed, we explore an alternative representation
of the phoretic velocities~(\ref{eq:V_Om_curlf}) introduced originally
in \rcite{DPRD20}, and which is valid in the quasi-homogeneous
regime. By using the expansions~(\ref{eq:alm}, \ref{eq:mulm}) and a
similar one for the adsorption potential,
\begin{equation}
  \label{eq:wlm}
  \mathbb{W}(\br=(R+z)\be_r) = \mathcal{W} \sum_{\ell m} w_{\ell m}(z)
  Y_{\ell m} (\be_r) ,
\end{equation}
where $\mathcal{W}$ is a characteristic scale of the potential and the
coefficients $w_{\ell m}(z)$ are dimensionless, one can write
(see App.~\ref{app:Yexp})
\begin{subequations}
  \label{eq:VOm_lm}
  \begin{equation}
    \label{eq:V_lm}
    \bV = V_0 \sum_{\ell m} \sum_{\ell' m'}
    a_{\ell' m'} \left\{
      \left[
        b_{\ell m; \ell'}
        + \frac{\ell'+1}{\ell+1}c_{\ell m; \ell'}
      \right]
      \,\bG^\parallel_{\ell m; \ell' m'}
      + c_{\ell m; \ell'} \, \bG^\parallel_{\ell' m'; \ell m}
    \right\} ,
  \end{equation}
  \begin{equation}
    \label{eq:Om_lm}
    \bOmega = \frac{3 V_0}{2 R}
    \sum_{\ell m} \sum_{\ell' m'}
    a_{\ell' m'} \left[
      b_{\ell m; \ell'-1}
      + \frac{1}{\ell+1}c_{\ell m; \ell'-1}
      \right] \bG^\tau_{\ell m; \ell' m'},
  \end{equation}
\end{subequations}
with
\begin{equation}
  \label{eq:V0}
  V_0 = \frac{\mathcal{A} \mathcal{W} R^2}{6\pi \eta \Gamma n_0 h''(n_0)} ,
\end{equation}
a characteristic velocity scale\footnote{We note that the
    expansions~(\ref{eq:VOm_lm}) are slightly simpler than the
    equivalent ones presented in \rcite{DPRD20} because we have
    applied two additional identities, see Eqs.~(\ref{eq:dlm}) and
    (\ref{eq:Gperp}).}. The dimensionless vectors
$\bG^{\parallel,\tau}$ (surface averages of products of spherical
harmonics defined in \rcite{DPRD20}) are given by
\begin{subequations}
  \label{eq:G}
  \begin{equation}
    \label{eq:Gpar}
    \bG^\parallel_{\ell m; \ell' m'} := - \frac{4 \pi R}{\ell'+1}
    \left\langle
      Y_{\ell m}(\be_r) \,  \nabla_\parallel Y_{\ell' m'}(\be_r)
    \right\rangle ,
  \end{equation}
  \begin{equation}
    \label{eq:Gtau}
    \bG^\tau_{\ell m; \ell' m'} := - \frac{4 \pi R}{\ell'+1}
    \left\langle
      Y_{\ell m}(\be_r) \,  \be_r\times\nabla_\parallel Y_{\ell' m'}(\be_r)
    \right\rangle ,
  \end{equation}
\end{subequations}
and the coefficients $b_{\ell m; \ell'}$ and $c_{\ell m; \ell'}$ are
defined as the following dimensionless radial integrals:
\begin{subequations}
  \label{eq:g_lm}
  \begin{equation}
    \label{eq:b_lm}
    b_{\ell m; \ell'} := \int\limits_{0}^\infty
    \frac{dz}{R} \;
    H^\mathrm{(b)}\left(\frac{z}{R}\right)
    \left( 1 + \frac{z}{R} \right)^{-\ell'} w_{\ell m}(z) ,
  \end{equation}
  \begin{equation}
    \label{eq:c_lm}
    c_{\ell m; \ell'} := -\frac{\ell+1}{2} \int\limits_{0}^\infty
    \frac{dz}{R} \;
    H^\mathrm{(c)}\left(\frac{z}{R}\right)
    \left( 1 + \frac{z}{R} \right)^{-\ell'} w_{\ell m}(z) ,
  \end{equation}
\end{subequations}
where the functions $H^{(b,c)}$, which are directly related to
the hydrodynamic kernels $A(r)$, $\Phi(r)$ in \eq{eq:V_Om_curlf},
are given by 
\begin{subequations}
  \label{eq:Khydro}
  \begin{equation}
    \label{eq:Kpar}
    H^\mathrm{(b)} (\zeta) := 1 -\frac{3}{4(1+\zeta)} -
    \frac{1}{4(1+\zeta)^3} ,
  \end{equation}
  \begin{equation}
    \label{eq:Kperp}
    H^\mathrm{(c)} (\zeta) := 1 -\frac{3}{2(1+\zeta)} +
    \frac{1}{2(1+\zeta)^3} .
  \end{equation}
\end{subequations}
One can recognize in the definitions~(\ref{eq:g_lm}) a
contribution ($w_{\ell m}$) coming from the the adsorption potential and a
$z$-dependent term stemming from the solution~(\ref{eq:mulm}) for the
chemical potential).

The most relevant feature concerning the $\bG$ vectors, which by their
definition are of the order of unity (when non-vanishing), are the so-called ``selection rules'' \cite{DPRD20}: in the double sum
in \eq{eq:V_lm}, the only terms different from zero are those which
satisfy $|\ell-\ell'|=1$ and $|m+m'|=0$ or $1$, while in \eq{eq:Om_lm}
only those which obey $|\ell-\ell'|=0$ and $|m+m'|=0$ or
$1$. Therefore, and against appearances, \eqs{eq:VOm_lm} involve a
single series over surface modes (because
$\ell -1 \leq \ell' \leq \ell+1$ and thus effectively the second
series is reduced to a few terms sum), each one characterized by a
length scale $\Lambda_\ell = R/\ell$ of lateral variation. This
property already provides an insightful observation: even if the
products involving $\mathbb{W}$ and $\mu$ in \eqs{eq:voms_lubr1} would
yield non-vanishing surface fields $\bv_s$, $\bomega_s$, their
contribution to the phoretic velocities is strongly suppressed upon
performing the surface average in \eq{eq:slips} if the fields
$\mathbb{W}$ and $\mathbb{A}$ have widely differing scales of lateral
variations, i.e., if their respective expansions are dominated by
  widely different values of $\ell$ and $\ell'$,
  respectively. Therefore, the length scales $\Lambda_\mathbb{W}$ and
$\Lambda_\mathbb{A}$ employed in the reasoning leading to the
lubrication approximation must effectively be of the same order. In
other words, it is sufficient to require one of the length scales,
e.g., $\Lambda_\mathbb{W}$, to be much larger than $\sigma$, and the
selection rules automatically ensure the proper lubrication
approximation in that a non-zero velocity can occur only if the other
length scale $\Lambda_\mathbb{A}$ is also large.

Assume once more that the coefficients $w_{\ell m}(R+z)$ vanish
quickly for separations $z \gtrsim \sigma$ from the surface. The
thin--layer approximation is recovered when $\sigma\ll R$, and is
expressed in \eqs{eq:VOm_lm} by Taylor-expanding the hydrodynamic
$H$-functions:
\begin{equation}
  \label{eq:Kthin}
  H^\mathrm{(b)} (\zeta \ll 1) \approx \frac{3\zeta}{2},
  \quad
  H^\mathrm{(c)} (\zeta \ll 1) \approx \frac{3\zeta^2}{2} .
\end{equation}
If, in addition, the main contribution to the sums in \eqs{eq:VOm_lm}
stems from modes with a small value of $\ell$, i.e., modes that vary
on the surface over a large scale
$\Lambda_\ell \sim R/\ell \sim R \gg \sigma$, one can expand
$(1 + z/R)^{-\ell'} \approx 1 - \ell' z/R = 1 - z/\Lambda_{\ell'}
\approx 1$ in the radial integrals. In this
case, the coefficient $c_{\ell m; \ell'}$ is a factor
$\sim (\ell+1) z/R \lesssim \sigma/R$ smaller than $b_{\ell m; \ell'}$, and thus the $c$ coefficients
can be neglected. In this manner, the lubrication approximation is recovered from \eqs{eq:VOm_lm} with the coefficient
\begin{equation}
  \label{eq:blubr}
  b^\mathrm{(lubr)}_{\ell m; \ell'} = \frac{3}{2 R^2}
  \int\limits_{0}^\infty
  dz \; z \, w_{\ell m}(z) ,
\end{equation}
compare with the phoretic coefficient~(\ref{eq:phoreticL}). The fact
that $b_{\ell m; \ell'}$ does not actually depend on the index $\ell'$
in this approximation reflects the factorization expressed by
\eq{eq:vs_lubr_quasihom}: in \eqs{eq:VOm_lm} the sum over $\ell'$ now involves only the activity, while that over $\ell$ only the adsorption
potential.

If, however, the sums in \eqs{eq:VOm_lm} get significant contributions
from modes with $\ell\gg 1$, associated to lateral variations over small scales $\Lambda_\ell \sim \sigma$, only the thin--layer
approximation~(\ref{eq:Kthin}) is expected to hold. In the radial
integrals one could then approximate
\begin{equation}
    \left( 1 + \frac{z}{R} \right)^{-\ell'} = \mathrm{e}^{-\ell' \ln
    (1+z/R)} \approx \mathrm{e}^{-\ell' z/R} =
  \mathrm{e}^{-z/\Lambda_{\ell'}} , 
\end{equation}
because $z/R \lesssim \sigma/R\ll 1$, and write
\begin{subequations}
  \label{eq:gthin}
  \begin{equation}
    \label{eq:bthin}
    b^\mathrm{(thin)}_{\ell m; \ell'} = \frac{3}{2R^2} \int\limits_{0}^\infty
    dz \;
    z \, \mathrm{e}^{-z/\Lambda_{\ell'}} \, w_{\ell m}(z) ,
  \end{equation}
  \begin{equation}
    \label{eq:cthin}
    c^\mathrm{(thin)}_{\ell m; \ell'} = - \frac{3}{4 R^2}
    \int\limits_{0}^\infty
    dz \;
    \frac{z^2}{\Lambda_\ell} \, 
    \mathrm{e}^{-z/\Lambda_{\ell'}} \, w_{\ell m}(z) ,
  \end{equation}
\end{subequations}
The relevant feature is the non-trivial dependence of the coefficients
on both indices $\ell, \ell'$, which prevents a simple factorization
as discussed above. Actually, there does not seem to be a simple and
appealing description, like the one based on the phoretic coefficient,
in the more general case.

\begin{figure}[t]
    \centering
    \hfill\includegraphics[width=0.45\textwidth]{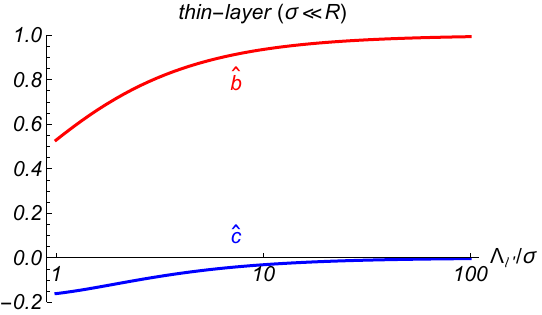}
    \hfill\includegraphics[width=0.45\textwidth]{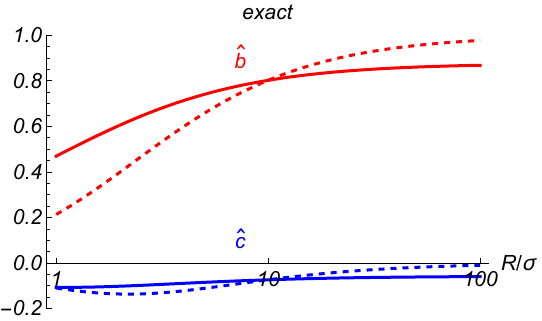}
    \hspace*{\fill}
    \caption{Plot of the ratios
      $\hat{b} := b_{\ell m;\ell'}/b^\mathrm{(lubr)}_{\ell
        m;\ell'}$ 
      (red curves) and
      $\hat{c}:=c_{\ell m;\ell'}/b^\mathrm{(lubr)}_{\ell
        m;\ell'}$ 
      (blue curves). The results do not depend on the index $m$, the
      dependence on $\ell'$ is parametrized in terms of the length
      $\Lambda_{\ell'}=R/\ell'$, and $\ell=\ell'$ is taken for
      simplicity in evaluating Eqs.~(\ref{eq:c_lm},
      \ref{eq:cthin_stepW}) for the coefficient $c_{\ell m;\ell'}$
      (since $|\ell-\ell'|\leq 1$ anyhow by the selection
      rules). \textbf{(Left panel}) The thin--layer approximation given by
      \eqs{eq:gthin_stepW}, as a function of
      $\Lambda_{\ell'}/\sigma$. In the limit
      $\Lambda_{\ell'}\gg\sigma$ one recovers the lubrication
      approximation~(\ref{eq:blubr}) (with a negligible value of the
      $c$-coefficient). 
            \textbf{(Right panel)} The exact values given by \eqs{eq:g_lm}:
      the dashed lines correspond to the choice $\Lambda_{\ell'}=R/2$
      (i.e., $\ell'=2$), the full lines to $\Lambda_{\ell'}=5
      \sigma$. In the limit $R\gg\sigma$ one recovers the lubrication
      approximation (dashed lines) or just the thin--layer
      approximation (full lines).}
    \label{fig:thinStepW}
  \end{figure}

Since the integrals appearing in \eqs{eq:gthin} do not seem much
different from \eq{eq:blubr} or \eqs{eq:g_lm}, the question of their
relative magnitude naturally arises. By way of example, we consider an adsorption potential modeled as a
square--well in the radial direction, so that each mode takes the form
($\hat{w}_{\ell m}$ is $z$-independent)
\begin{equation}
  \label{eq:stepW}
  w_{\ell m}(R+z) = \left\{
    \begin{array}[c]{cl}
      \hat{w}_{\ell m} ,
      & 0< z < \sigma ,
      \\
      0 ,
      & \sigma < z .
    \end{array}
  \right.
\end{equation}
Then, the lubrication approximation~(\ref{eq:blubr}) gives
\begin{equation}
  \label{eq:blubr_stepW}
  b^\mathrm{(lubr)}_{\ell m; \ell'} = \frac{3 \sigma^2}{4 R^2} \hat{w}_{\ell m} ,
\end{equation}
and the thin--layer approximation~(\ref{eq:gthin}) yields
\begin{subequations}
  \label{eq:gthin_stepW}
  \begin{equation}
    b^\mathrm{(thin)}_{\ell m; \ell'} =
    b^\mathrm{(lubr)}_{\ell m; \ell'}
    \; \frac{2\Lambda_{\ell'}^2}{\sigma^2} \left[
      1 - \left(1 + \frac{\sigma}{\Lambda_{\ell'}}\right)
      \mathrm{e}^{-\sigma/\Lambda_{\ell'}}
    \right] ,
  \end{equation}
  \begin{equation}
    \label{eq:cthin_stepW}
    c^\mathrm{(thin)}_{\ell m; \ell'} =
    - \frac{\ell}{\ell'} b^\mathrm{(lubr)}_{\ell m; \ell'}
    \; \frac{2\Lambda_{\ell'}^2}{\sigma^2} \left[
      1 - \left(1 + \frac{\sigma}{\Lambda_{\ell'}}
        + \frac{\sigma^2}{2\Lambda_{\ell'}^2} \right)
      \mathrm{e}^{-\sigma/\Lambda_{\ell'}}
    \right] .
  \end{equation}
\end{subequations}
Figure~\ref{fig:thinStepW}(left) compares the coefficients in the
lubrication or in the thin--layer approximations, respectively, as a
function of the ratio $\Lambda_{\ell'}/\sigma$ (with the choice
$\ell' = \ell$, since the selection rules imply that they must have
similar values to contribute to the
velocity). One can straightforwardly infer that the changes are rather
mild even at small values of $\Lambda_{\ell'}$; thus one concludes
that the results of the lubrication approximation may provide
reasonable estimates for the velocity even in conditions under which
the underlying approximations are not expected to hold.

One can also compare with the exact results that follow from
\eqs{eq:g_lm}, i.e., before any assumption is made on the relative
magnitude of the adsorption range $\sigma$ and the radius
$R$. Although the integrals can be evaluated exactly for the
choice~(\ref{eq:stepW}), the results are lengthy and not very
illuminating, and we do not deem necessary to present them. For
illustration, Fig.~\ref{fig:thinStepW}(right) shows the comparison of
the exact results with the lubrication approximation as a function of
the ratio $R/\sigma$ in two representative cases, namely, when the
scale of lateral variation verifies $\Lambda_{\ell'} \sim R$, and when
it is $\Lambda_{\ell'} \sim \sigma$, respectively. The conclusion that
follows from Fig.~(\ref{fig:thinStepW}) is that there are not major
quantitative changes when different approximations are considered
provided the range $\sigma$ and the scales of lateral variation are
smaller or comparable to the particle radius.

  It thus seems that, as far as the quantitative value of the radial
  coefficients~(\ref{eq:g_lm}) is concerned, one could use the
  lubrication approximation in order to get good estimates even beyond
  its expected range of validity. Although this insight is based on the specific choice of a
  square--well adsorption potential~(\ref{eq:stepW}), one can arguably
  expect it to hold for more generic forms that have a single
  characteristic scale $\sigma$ of smooth variation.

\section{Conclusions}

By using the case of a spherical active particle, whose activity
consists in releasing or removing a solute and moves by
self-chemophoresis, we re-examined the analysis of the rigid-body
motion under the assumption of a thin diffuse-layer by employing an
integral representation. This allowed us to decouple two steps of
approximation: the thin diffuse interface (\textit{``thin layer''})
and the lubrication (slow variations along the tangential direction)
for the adsorption potential, which in general have been implicitly
combined by calculations in a planar geometry. Consequently, this
permits a straightforward, systematic analysis of the motion in each
of the two regimes of interest defined by whether or not the
lubrication approximation for the adsorption potential holds.

The lubrication approximation means that the range of the adsorption
potential is much smaller larger than any other relevant length scale
of the model. In this approximation and under the assumption of small
deviations from the equilibrium state
(\textit{``quasi--homogeneity''}), the approach provides a very
simple, straightforward derivation of the classic expressions for a
phoretic slip velocity in terms of a phoretic coefficient. A potential
line of further research consists in relaxing the hypothesis of
quasi--homogeneity: this has been done so far only for an ideal gas,
which leads to the classical non-linear dependence of the phoretic
coefficient on the adsorption potential. It seems worth exploring how
this latter result is affected by solute--solute interactions.

One can also relax the lubrication approximation by allowing
variations of the adsorption potential over the surface on a length
scale comparable with the thickness of the diffuse interface. This
case has been addressed explicitly in this work and our analysis
suggests that the phoretic velocities would not change much
quantitatively. But then, this prompts the study of the opposite
limit, namely, when the range of the adsorption potential is
significantly larger than any other length scale, particularly,
  the particle radius. 

We expect that the analysis and the results presented in this paper
will provide a further impetus to the theoretical interest in the
issue of self-phoresis, as well as motivation for experimental studies
focused on active particles with patterned surfaces engineered at
small scales, both for a validation of the theoretical predictions
and, more importantly, as a step towards investigating the intriguing
case of ``no diffuse interface'' mentioned above.

\section*{Acknowledgments and dedication}

The authors acknowledge financial support through grants
ProyExcel\_00505 funded by Junta de Andaluc{\'i}a, and
PID2021-126348NB-I00 funded by MCIN/AEI/10.13039/501100011033 and
``ERDF A way of making Europe''. MNP~also acknowledges support from
Ministerio de Universidades through a Mar{\'i}a Zambrano grant. AD dedicates this manuscript to the late Prof.~Luis F.~Rull, a dear mentor who introduced AD into scientific research.


\begin{thebibliography}{49}%
\makeatletter
\providecommand \@ifxundefined [1]{%
 \@ifx{#1\undefined}
}%
\providecommand \@ifnum [1]{%
 \ifnum #1\expandafter \@firstoftwo
 \else \expandafter \@secondoftwo
 \fi
}%
\providecommand \@ifx [1]{%
 \ifx #1\expandafter \@firstoftwo
 \else \expandafter \@secondoftwo
 \fi
}%
\providecommand \natexlab [1]{#1}%
\providecommand \enquote  [1]{``#1''}%
\providecommand \bibnamefont  [1]{#1}%
\providecommand \bibfnamefont [1]{#1}%
\providecommand \citenamefont [1]{#1}%
\providecommand \href@noop [0]{\@secondoftwo}%
\providecommand \href [0]{\begingroup \@sanitize@url \@href}%
\providecommand \@href[1]{\@@startlink{#1}\@@href}%
\providecommand \@@href[1]{\endgroup#1\@@endlink}%
\providecommand \@sanitize@url [0]{\catcode `\\12\catcode `\$12\catcode
  `\&12\catcode `\#12\catcode `\^12\catcode `\_12\catcode `\%12\relax}%
\providecommand \@@startlink[1]{}%
\providecommand \@@endlink[0]{}%
\providecommand \url  [0]{\begingroup\@sanitize@url \@url }%
\providecommand \@url [1]{\endgroup\@href {#1}{\urlprefix }}%
\providecommand \urlprefix  [0]{URL }%
\providecommand \Eprint [0]{\href }%
\providecommand \doibase [0]{https://doi.org/}%
\providecommand \selectlanguage [0]{\@gobble}%
\providecommand \bibinfo  [0]{\@secondoftwo}%
\providecommand \bibfield  [0]{\@secondoftwo}%
\providecommand \translation [1]{[#1]}%
\providecommand \BibitemOpen [0]{}%
\providecommand \bibitemStop [0]{}%
\providecommand \bibitemNoStop [0]{.\EOS\space}%
\providecommand \EOS [0]{\spacefactor3000\relax}%
\providecommand \BibitemShut  [1]{\csname bibitem#1\endcsname}%
\let\auto@bib@innerbib\@empty
\bibitem [{\citenamefont {Ebbens}\ and\ \citenamefont
  {Howse}(2010)}]{Ebbens2010}%
  \BibitemOpen
  \bibfield  {author} {\bibinfo {author} {\bibfnamefont {S.~J.}\ \bibnamefont
  {Ebbens}}\ and\ \bibinfo {author} {\bibfnamefont {J.~R.}\ \bibnamefont
  {Howse}},\ }\bibfield  {title} {\bibinfo {title} {In pursuit of propulsion at
  the nanoscale},\ }\href@noop {} {\bibfield  {journal} {\bibinfo  {journal}
  {Soft Matter}\ }\textbf {\bibinfo {volume} {6}},\ \bibinfo {pages} {726}
  (\bibinfo {year} {2010})}\BibitemShut {NoStop}%
\bibitem [{\citenamefont {Bechinger}\ \emph {et~al.}(2016)\citenamefont
  {Bechinger}, \citenamefont {Leonardo}, \citenamefont {L{\"o}wen},
  \citenamefont {Reichhardt}, \citenamefont {Volpe},\ and\ \citenamefont
  {Volpe}}]{BDLR16}%
  \BibitemOpen
  \bibfield  {author} {\bibinfo {author} {\bibfnamefont {C.}~\bibnamefont
  {Bechinger}}, \bibinfo {author} {\bibfnamefont {R.~D.}\ \bibnamefont
  {Leonardo}}, \bibinfo {author} {\bibfnamefont {H.}~\bibnamefont {L{\"o}wen}},
  \bibinfo {author} {\bibfnamefont {C.}~\bibnamefont {Reichhardt}}, \bibinfo
  {author} {\bibfnamefont {G.}~\bibnamefont {Volpe}},\ and\ \bibinfo {author}
  {\bibfnamefont {G.}~\bibnamefont {Volpe}},\ }\bibfield  {title} {\bibinfo
  {title} {Active particles in complex and crowded environments},\ }\href@noop
  {} {\bibfield  {journal} {\bibinfo  {journal} {Rev. Mod. Phys.}\ }\textbf
  {\bibinfo {volume} {88}},\ \bibinfo {pages} {045006} (\bibinfo {year}
  {2016})}\BibitemShut {NoStop}%
\bibitem [{\citenamefont {Dey}\ \emph {et~al.}(2016)\citenamefont {Dey},
  \citenamefont {Wong}, \citenamefont {Altemose},\ and\ \citenamefont
  {Sen}}]{DEY2016}%
  \BibitemOpen
  \bibfield  {author} {\bibinfo {author} {\bibfnamefont {K.~K.}\ \bibnamefont
  {Dey}}, \bibinfo {author} {\bibfnamefont {F.}~\bibnamefont {Wong}}, \bibinfo
  {author} {\bibfnamefont {A.}~\bibnamefont {Altemose}},\ and\ \bibinfo
  {author} {\bibfnamefont {A.}~\bibnamefont {Sen}},\ }\bibfield  {title}
  {\bibinfo {title} {{Catalytic motors—Quo Vadimus?}},\ }\href@noop {}
  {\bibfield  {journal} {\bibinfo  {journal} {Curr. Op. Colloid Interf. Sci.}\
  }\textbf {\bibinfo {volume} {21}},\ \bibinfo {pages} {4 } (\bibinfo {year}
  {2016})}\BibitemShut {NoStop}%
\bibitem [{\citenamefont {Gompper}\ \emph {et~al.}(2020)\citenamefont
  {Gompper}, \citenamefont {Winkler}, \citenamefont {Speck}, \citenamefont
  {Solon}, \citenamefont {Nardini}, \citenamefont {Peruani}, \citenamefont
  {L{\"o}wen}, \citenamefont {Golestanian}, \citenamefont {Kaupp},
  \citenamefont {Alvarez}, \citenamefont {Ki{\/o}rboe}, \citenamefont {Lauga},
  \citenamefont {Poon}, \citenamefont {DeSimone}, \citenamefont
  {Mui{\~n}os-Landin}, \citenamefont {Fischer}, \citenamefont {S{\"o}ker},
  \citenamefont {Cichos}, \citenamefont {Kapral}, \citenamefont {Gaspard},
  \citenamefont {Ripoll}, \citenamefont {Sagues}, \citenamefont
  {Doostmohammadi}, \citenamefont {Yeomans}, \citenamefont {Aranson},
  \citenamefont {Bechinger}, \citenamefont {Stark}, \citenamefont {Hemelrijk},
  \citenamefont {Nedelec}, \citenamefont {Sarkar}, \citenamefont {Aryaksama},
  \citenamefont {Lacroix}, \citenamefont {Duclos}, \citenamefont {Yashunsky},
  \citenamefont {Silberzan}, \citenamefont {Arroyo},\ and\ \citenamefont
  {Kale}}]{GWSS20}%
  \BibitemOpen
  \bibfield  {author} {\bibinfo {author} {\bibfnamefont {G.}~\bibnamefont
  {Gompper}}, \bibinfo {author} {\bibfnamefont {R.~G.}\ \bibnamefont
  {Winkler}}, \bibinfo {author} {\bibfnamefont {T.}~\bibnamefont {Speck}},
  \bibinfo {author} {\bibfnamefont {A.}~\bibnamefont {Solon}}, \bibinfo
  {author} {\bibfnamefont {C.}~\bibnamefont {Nardini}}, \bibinfo {author}
  {\bibfnamefont {F.}~\bibnamefont {Peruani}}, \bibinfo {author} {\bibfnamefont
  {H.}~\bibnamefont {L{\"o}wen}}, \bibinfo {author} {\bibfnamefont
  {R.}~\bibnamefont {Golestanian}}, \bibinfo {author} {\bibfnamefont {U.~B.}\
  \bibnamefont {Kaupp}}, \bibinfo {author} {\bibfnamefont {L.}~\bibnamefont
  {Alvarez}}, \bibinfo {author} {\bibfnamefont {T.}~\bibnamefont
  {Ki{\/o}rboe}}, \bibinfo {author} {\bibfnamefont {E.}~\bibnamefont {Lauga}},
  \bibinfo {author} {\bibfnamefont {W.~C.~K.}\ \bibnamefont {Poon}}, \bibinfo
  {author} {\bibfnamefont {A.}~\bibnamefont {DeSimone}}, \bibinfo {author}
  {\bibfnamefont {S.}~\bibnamefont {Mui{\~n}os-Landin}}, \bibinfo {author}
  {\bibfnamefont {A.}~\bibnamefont {Fischer}}, \bibinfo {author} {\bibfnamefont
  {N.~A.}\ \bibnamefont {S{\"o}ker}}, \bibinfo {author} {\bibfnamefont
  {F.}~\bibnamefont {Cichos}}, \bibinfo {author} {\bibfnamefont
  {R.}~\bibnamefont {Kapral}}, \bibinfo {author} {\bibfnamefont
  {P.}~\bibnamefont {Gaspard}}, \bibinfo {author} {\bibfnamefont
  {M.}~\bibnamefont {Ripoll}}, \bibinfo {author} {\bibfnamefont
  {F.}~\bibnamefont {Sagues}}, \bibinfo {author} {\bibfnamefont
  {A.}~\bibnamefont {Doostmohammadi}}, \bibinfo {author} {\bibfnamefont
  {J.~M.}\ \bibnamefont {Yeomans}}, \bibinfo {author} {\bibfnamefont {I.~S.}\
  \bibnamefont {Aranson}}, \bibinfo {author} {\bibfnamefont {C.}~\bibnamefont
  {Bechinger}}, \bibinfo {author} {\bibfnamefont {H.}~\bibnamefont {Stark}},
  \bibinfo {author} {\bibfnamefont {C.~K.}\ \bibnamefont {Hemelrijk}}, \bibinfo
  {author} {\bibfnamefont {F.~J.}\ \bibnamefont {Nedelec}}, \bibinfo {author}
  {\bibfnamefont {T.}~\bibnamefont {Sarkar}}, \bibinfo {author} {\bibfnamefont
  {T.}~\bibnamefont {Aryaksama}}, \bibinfo {author} {\bibfnamefont
  {M.}~\bibnamefont {Lacroix}}, \bibinfo {author} {\bibfnamefont
  {G.}~\bibnamefont {Duclos}}, \bibinfo {author} {\bibfnamefont
  {V.}~\bibnamefont {Yashunsky}}, \bibinfo {author} {\bibfnamefont
  {P.}~\bibnamefont {Silberzan}}, \bibinfo {author} {\bibfnamefont
  {M.}~\bibnamefont {Arroyo}},\ and\ \bibinfo {author} {\bibfnamefont
  {S.}~\bibnamefont {Kale}},\ }\bibfield  {title} {\bibinfo {title} {The 2020
  motile active matter roadmap},\ }\href@noop {} {\bibfield  {journal}
  {\bibinfo  {journal} {J. Phys. Condens. Matt.}\ }\textbf {\bibinfo {volume}
  {32}},\ \bibinfo {pages} {193001} (\bibinfo {year} {2020})}\BibitemShut
  {NoStop}%
\bibitem [{\citenamefont {Palacci}\ \emph {et~al.}(2013)\citenamefont
  {Palacci}, \citenamefont {Sacanna}, \citenamefont {Steinberg}, \citenamefont
  {Pine},\ and\ \citenamefont {Chaikin}}]{Palacci2013}%
  \BibitemOpen
  \bibfield  {author} {\bibinfo {author} {\bibfnamefont {J.}~\bibnamefont
  {Palacci}}, \bibinfo {author} {\bibfnamefont {S.}~\bibnamefont {Sacanna}},
  \bibinfo {author} {\bibfnamefont {A.~P.}\ \bibnamefont {Steinberg}}, \bibinfo
  {author} {\bibfnamefont {D.~J.}\ \bibnamefont {Pine}},\ and\ \bibinfo
  {author} {\bibfnamefont {P.~M.}\ \bibnamefont {Chaikin}},\ }\bibfield
  {title} {\bibinfo {title} {Living crystals of light-activated colloidal
  surfers},\ }\href@noop {} {\bibfield  {journal} {\bibinfo  {journal}
  {Science}\ }\textbf {\bibinfo {volume} {339}},\ \bibinfo {pages} {936 }
  (\bibinfo {year} {2013})}\BibitemShut {NoStop}%
\bibitem [{\citenamefont {Aubret}\ \emph {et~al.}(2018)\citenamefont {Aubret},
  \citenamefont {Youssef}, \citenamefont {Sacanna},\ and\ \citenamefont
  {Palacci}}]{Palacci2018}%
  \BibitemOpen
  \bibfield  {author} {\bibinfo {author} {\bibfnamefont {A.}~\bibnamefont
  {Aubret}}, \bibinfo {author} {\bibfnamefont {M.}~\bibnamefont {Youssef}},
  \bibinfo {author} {\bibfnamefont {S.}~\bibnamefont {Sacanna}},\ and\ \bibinfo
  {author} {\bibfnamefont {J.}~\bibnamefont {Palacci}},\ }\bibfield  {title}
  {\bibinfo {title} {Targeted assembly and synchronization of self-spinning
  microgears},\ }\href@noop {} {\bibfield  {journal} {\bibinfo  {journal}
  {Nature Phys.}\ }\textbf {\bibinfo {volume} {14}},\ \bibinfo {pages} {1114}
  (\bibinfo {year} {2018})}\BibitemShut {NoStop}%
\bibitem [{\citenamefont {Maggi}\ \emph {et~al.}(2016)\citenamefont {Maggi},
  \citenamefont {Simmchen}, \citenamefont {Saglimbeni}, \citenamefont {Katuri},
  \citenamefont {Dipalo}, \citenamefont {De~Angelis}, \citenamefont {Sanchez},\
  and\ \citenamefont {Di~Leonardo}}]{DiLeonardo2016}%
  \BibitemOpen
  \bibfield  {author} {\bibinfo {author} {\bibfnamefont {C.}~\bibnamefont
  {Maggi}}, \bibinfo {author} {\bibfnamefont {J.}~\bibnamefont {Simmchen}},
  \bibinfo {author} {\bibfnamefont {F.}~\bibnamefont {Saglimbeni}}, \bibinfo
  {author} {\bibfnamefont {J.}~\bibnamefont {Katuri}}, \bibinfo {author}
  {\bibfnamefont {M.}~\bibnamefont {Dipalo}}, \bibinfo {author} {\bibfnamefont
  {F.}~\bibnamefont {De~Angelis}}, \bibinfo {author} {\bibfnamefont
  {S.}~\bibnamefont {Sanchez}},\ and\ \bibinfo {author} {\bibfnamefont
  {R.}~\bibnamefont {Di~Leonardo}},\ }\bibfield  {title} {\bibinfo {title}
  {Self-assembly of micromachining systems powered by {Janus} micromotors},\
  }\href@noop {} {\bibfield  {journal} {\bibinfo  {journal} {Small}\ }\textbf
  {\bibinfo {volume} {12}},\ \bibinfo {pages} {446} (\bibinfo {year}
  {2016})}\BibitemShut {NoStop}%
\bibitem [{\citenamefont {Sundararajan}\ \emph {et~al.}(2008)\citenamefont
  {Sundararajan}, \citenamefont {Lammert}, \citenamefont {Zudans},
  \citenamefont {Crespi},\ and\ \citenamefont {Sen}}]{Sen2008}%
  \BibitemOpen
  \bibfield  {author} {\bibinfo {author} {\bibfnamefont {S.}~\bibnamefont
  {Sundararajan}}, \bibinfo {author} {\bibfnamefont {P.~E.}\ \bibnamefont
  {Lammert}}, \bibinfo {author} {\bibfnamefont {A.~W.}\ \bibnamefont {Zudans}},
  \bibinfo {author} {\bibfnamefont {V.~H.}\ \bibnamefont {Crespi}},\ and\
  \bibinfo {author} {\bibfnamefont {A.}~\bibnamefont {Sen}},\ }\bibfield
  {title} {\bibinfo {title} {Catalytic motors for transport of colloidal
  cargo},\ }\href@noop {} {\bibfield  {journal} {\bibinfo  {journal} {Nano
  Letters}\ }\textbf {\bibinfo {volume} {8}},\ \bibinfo {pages} {1271}
  (\bibinfo {year} {2008})}\BibitemShut {NoStop}%
\bibitem [{\citenamefont {Baraban}\ \emph {et~al.}(2012)\citenamefont
  {Baraban}, \citenamefont {Tasinkevych}, \citenamefont {Popescu},
  \citenamefont {Sanchez}, \citenamefont {Dietrich},\ and\ \citenamefont
  {Schmidt}}]{Baraban2012}%
  \BibitemOpen
  \bibfield  {author} {\bibinfo {author} {\bibfnamefont {L.}~\bibnamefont
  {Baraban}}, \bibinfo {author} {\bibfnamefont {M.}~\bibnamefont
  {Tasinkevych}}, \bibinfo {author} {\bibfnamefont {M.~N.}\ \bibnamefont
  {Popescu}}, \bibinfo {author} {\bibfnamefont {S.}~\bibnamefont {Sanchez}},
  \bibinfo {author} {\bibfnamefont {S.}~\bibnamefont {Dietrich}},\ and\
  \bibinfo {author} {\bibfnamefont {O.}~\bibnamefont {Schmidt}},\ }\bibfield
  {title} {\bibinfo {title} {Transport of cargo by catalytic {Janus}
  micro-motors},\ }\href@noop {} {\bibfield  {journal} {\bibinfo  {journal}
  {Soft Matter}\ }\textbf {\bibinfo {volume} {8}},\ \bibinfo {pages} {48}
  (\bibinfo {year} {2012})}\BibitemShut {NoStop}%
\bibitem [{\citenamefont {Duan}\ \emph {et~al.}(2015)\citenamefont {Duan},
  \citenamefont {Wang}, \citenamefont {Das}, \citenamefont {Yadav},
  \citenamefont {Mallouk},\ and\ \citenamefont {Sen}}]{Wang2015}%
  \BibitemOpen
  \bibfield  {author} {\bibinfo {author} {\bibfnamefont {W.}~\bibnamefont
  {Duan}}, \bibinfo {author} {\bibfnamefont {W.}~\bibnamefont {Wang}}, \bibinfo
  {author} {\bibfnamefont {S.}~\bibnamefont {Das}}, \bibinfo {author}
  {\bibfnamefont {V.}~\bibnamefont {Yadav}}, \bibinfo {author} {\bibfnamefont
  {T.~E.}\ \bibnamefont {Mallouk}},\ and\ \bibinfo {author} {\bibfnamefont
  {A.}~\bibnamefont {Sen}},\ }\bibfield  {title} {\bibinfo {title} {Synthetic
  nano- and micromachines in analytical chemistry: Sensing, migration, capture,
  delivery, and separation},\ }\href@noop {} {\bibfield  {journal} {\bibinfo
  {journal} {Annu. Rev. Analyt. Chem.}\ }\textbf {\bibinfo {volume} {8}},\
  \bibinfo {pages} {311} (\bibinfo {year} {2015})}\BibitemShut {NoStop}%
\bibitem [{\citenamefont {Katuri}\ \emph {et~al.}(2018)\citenamefont {Katuri},
  \citenamefont {Caballero}, \citenamefont {Voituriez}, \citenamefont
  {Samitier},\ and\ \citenamefont {S{\'a}nchez}}]{Katuri2018}%
  \BibitemOpen
  \bibfield  {author} {\bibinfo {author} {\bibfnamefont {J.}~\bibnamefont
  {Katuri}}, \bibinfo {author} {\bibfnamefont {D.}~\bibnamefont {Caballero}},
  \bibinfo {author} {\bibfnamefont {R.}~\bibnamefont {Voituriez}}, \bibinfo
  {author} {\bibfnamefont {J.}~\bibnamefont {Samitier}},\ and\ \bibinfo
  {author} {\bibfnamefont {S.}~\bibnamefont {S{\'a}nchez}},\ }\bibfield
  {title} {\bibinfo {title} {Directed flow of micromotors through alignment
  interactions with micropatterned ratchets},\ }\href@noop {} {\bibfield
  {journal} {\bibinfo  {journal} {ACS Nano}\ }\textbf {\bibinfo {volume}
  {12}},\ \bibinfo {pages} {7282} (\bibinfo {year} {2018})}\BibitemShut
  {NoStop}%
\bibitem [{\citenamefont {Bekir}\ \emph {et~al.}(2023)\citenamefont {Bekir},
  \citenamefont {Sperling}, \citenamefont {Mu{\~n}oz}, \citenamefont {Braksch},
  \citenamefont {B{\"o}ker}, \citenamefont {Lomadze}, \citenamefont {Popescu},\
  and\ \citenamefont {Santer}}]{Bekir2023}%
  \BibitemOpen
  \bibfield  {author} {\bibinfo {author} {\bibfnamefont {M.}~\bibnamefont
  {Bekir}}, \bibinfo {author} {\bibfnamefont {M.}~\bibnamefont {Sperling}},
  \bibinfo {author} {\bibfnamefont {D.~V.}\ \bibnamefont {Mu{\~n}oz}}, \bibinfo
  {author} {\bibfnamefont {C.}~\bibnamefont {Braksch}}, \bibinfo {author}
  {\bibfnamefont {A.}~\bibnamefont {B{\"o}ker}}, \bibinfo {author}
  {\bibfnamefont {N.}~\bibnamefont {Lomadze}}, \bibinfo {author} {\bibfnamefont
  {M.~N.}\ \bibnamefont {Popescu}},\ and\ \bibinfo {author} {\bibfnamefont
  {S.}~\bibnamefont {Santer}},\ }\bibfield  {title} {\bibinfo {title}
  {Versatile microfluidics separation of colloids by combining external flow
  with light-induced chemical activity},\ }\href@noop {} {\bibfield  {journal}
  {\bibinfo  {journal} {Advanced Materials}\ }\textbf {\bibinfo {volume}
  {35}},\ \bibinfo {pages} {2300358} (\bibinfo {year} {2023})}\BibitemShut
  {NoStop}%
\bibitem [{\citenamefont {Paxton}\ \emph {et~al.}(2004)\citenamefont {Paxton},
  \citenamefont {Kistler}, \citenamefont {Olmeda}, \citenamefont {Sen},
  \citenamefont {St.~Angelo}, \citenamefont {Cao}, \citenamefont {Mallouk},
  \citenamefont {Lammert},\ and\ \citenamefont {Crespi}}]{Paxton2004}%
  \BibitemOpen
  \bibfield  {author} {\bibinfo {author} {\bibfnamefont {W.~F.}\ \bibnamefont
  {Paxton}}, \bibinfo {author} {\bibfnamefont {K.~C.}\ \bibnamefont {Kistler}},
  \bibinfo {author} {\bibfnamefont {C.~C.}\ \bibnamefont {Olmeda}}, \bibinfo
  {author} {\bibfnamefont {A.}~\bibnamefont {Sen}}, \bibinfo {author}
  {\bibfnamefont {S.~K.}\ \bibnamefont {St.~Angelo}}, \bibinfo {author}
  {\bibfnamefont {Y.~Y.}\ \bibnamefont {Cao}}, \bibinfo {author} {\bibfnamefont
  {T.~E.}\ \bibnamefont {Mallouk}}, \bibinfo {author} {\bibfnamefont {P.~E.}\
  \bibnamefont {Lammert}},\ and\ \bibinfo {author} {\bibfnamefont {V.~H.}\
  \bibnamefont {Crespi}},\ }\bibfield  {title} {\bibinfo {title} {Catalytic
  nanomotors: Autonomous movement of striped nanorods},\ }\href@noop {}
  {\bibfield  {journal} {\bibinfo  {journal} {J. Am. Chem. Soc.}\ }\textbf
  {\bibinfo {volume} {126}},\ \bibinfo {pages} {13424} (\bibinfo {year}
  {2004})}\BibitemShut {NoStop}%
\bibitem [{\citenamefont {Fournier-Bidoz}\ \emph {et~al.}(2005)\citenamefont
  {Fournier-Bidoz}, \citenamefont {Arsenault}, \citenamefont {Manners},\ and\
  \citenamefont {Ozin}}]{Fournier-Bidoz2005}%
  \BibitemOpen
  \bibfield  {author} {\bibinfo {author} {\bibfnamefont {S.}~\bibnamefont
  {Fournier-Bidoz}}, \bibinfo {author} {\bibfnamefont {A.~C.}\ \bibnamefont
  {Arsenault}}, \bibinfo {author} {\bibfnamefont {I.}~\bibnamefont {Manners}},\
  and\ \bibinfo {author} {\bibfnamefont {G.~A.}\ \bibnamefont {Ozin}},\
  }\bibfield  {title} {\bibinfo {title} {Synthetic self-propelled nanorotors},\
  }\href@noop {} {\bibfield  {journal} {\bibinfo  {journal} {Chem. Commun.}\
  }\textbf {\bibinfo {volume} {0}},\ \bibinfo {pages} {441} (\bibinfo {year}
  {2005})}\BibitemShut {NoStop}%
\bibitem [{\citenamefont {Golestanian}\ \emph {et~al.}(2005)\citenamefont
  {Golestanian}, \citenamefont {Liverpool},\ and\ \citenamefont
  {Ajdari}}]{GLA05}%
  \BibitemOpen
  \bibfield  {author} {\bibinfo {author} {\bibfnamefont {R.}~\bibnamefont
  {Golestanian}}, \bibinfo {author} {\bibfnamefont {T.~B.}\ \bibnamefont
  {Liverpool}},\ and\ \bibinfo {author} {\bibfnamefont {A.}~\bibnamefont
  {Ajdari}},\ }\bibfield  {title} {\bibinfo {title} {Propulsion of a molecular
  machine by asymmetric distribution of reaction products},\ }\href@noop {}
  {\bibfield  {journal} {\bibinfo  {journal} {Phys. Rev. Lett.}\ }\textbf
  {\bibinfo {volume} {94}},\ \bibinfo {pages} {220801} (\bibinfo {year}
  {2005})}\BibitemShut {NoStop}%
\bibitem [{\citenamefont {Simmchen}\ \emph {et~al.}(2016)\citenamefont
  {Simmchen}, \citenamefont {Katuri}, \citenamefont {Uspal}, \citenamefont
  {Popescu}, \citenamefont {Tasinkevych},\ and\ \citenamefont
  {S{\'a}nchez}}]{Simmchen2016}%
  \BibitemOpen
  \bibfield  {author} {\bibinfo {author} {\bibfnamefont {J.}~\bibnamefont
  {Simmchen}}, \bibinfo {author} {\bibfnamefont {J.}~\bibnamefont {Katuri}},
  \bibinfo {author} {\bibfnamefont {W.~E.}\ \bibnamefont {Uspal}}, \bibinfo
  {author} {\bibfnamefont {M.~N.}\ \bibnamefont {Popescu}}, \bibinfo {author}
  {\bibfnamefont {M.}~\bibnamefont {Tasinkevych}},\ and\ \bibinfo {author}
  {\bibfnamefont {S.}~\bibnamefont {S{\'a}nchez}},\ }\bibfield  {title}
  {\bibinfo {title} {Topographical pathways guide chemical microswimmers},\
  }\href@noop {} {\bibfield  {journal} {\bibinfo  {journal} {Nat. Comm.}\
  }\textbf {\bibinfo {volume} {7}},\ \bibinfo {pages} {10598:1} (\bibinfo
  {year} {2016})}\BibitemShut {NoStop}%
\bibitem [{\citenamefont {Volpe}\ \emph {et~al.}(2011)\citenamefont {Volpe},
  \citenamefont {Buttinoni}, \citenamefont {Vogt}, \citenamefont
  {K{\"u}mmerer},\ and\ \citenamefont {Bechinger}}]{Volpe2011}%
  \BibitemOpen
  \bibfield  {author} {\bibinfo {author} {\bibfnamefont {G.}~\bibnamefont
  {Volpe}}, \bibinfo {author} {\bibfnamefont {I.}~\bibnamefont {Buttinoni}},
  \bibinfo {author} {\bibfnamefont {D.}~\bibnamefont {Vogt}}, \bibinfo {author}
  {\bibfnamefont {H.-J.}\ \bibnamefont {K{\"u}mmerer}},\ and\ \bibinfo {author}
  {\bibfnamefont {C.}~\bibnamefont {Bechinger}},\ }\bibfield  {title} {\bibinfo
  {title} {Microswimmers in patterned environments},\ }\href@noop {} {\bibfield
   {journal} {\bibinfo  {journal} {Soft Matter}\ }\textbf {\bibinfo {volume}
  {7}},\ \bibinfo {pages} {8810} (\bibinfo {year} {2011})}\BibitemShut
  {NoStop}%
\bibitem [{\citenamefont {Kroy}\ \emph {et~al.}(2016)\citenamefont {Kroy},
  \citenamefont {Chakraborty},\ and\ \citenamefont {Cichos}}]{Kroy2016}%
  \BibitemOpen
  \bibfield  {author} {\bibinfo {author} {\bibfnamefont {K.}~\bibnamefont
  {Kroy}}, \bibinfo {author} {\bibfnamefont {D.}~\bibnamefont {Chakraborty}},\
  and\ \bibinfo {author} {\bibfnamefont {F.}~\bibnamefont {Cichos}},\
  }\bibfield  {title} {\bibinfo {title} {Hot microswimmers},\ }\href@noop {}
  {\bibfield  {journal} {\bibinfo  {journal} {Eur. Phys. J. Spec. Topics}\
  }\textbf {\bibinfo {volume} {225}},\ \bibinfo {pages} {2207 } (\bibinfo
  {year} {2016})}\BibitemShut {NoStop}%
\bibitem [{\citenamefont {Derjaguin}\ \emph {et~al.}(1947)\citenamefont
  {Derjaguin}, \citenamefont {Sidorenkov}, \citenamefont {Zubashchenkov},\ and\
  \citenamefont {Kiseleva}}]{DSZK47}%
  \BibitemOpen
  \bibfield  {author} {\bibinfo {author} {\bibfnamefont {B.~V.}\ \bibnamefont
  {Derjaguin}}, \bibinfo {author} {\bibfnamefont {G.}~\bibnamefont
  {Sidorenkov}}, \bibinfo {author} {\bibfnamefont {E.}~\bibnamefont
  {Zubashchenkov}},\ and\ \bibinfo {author} {\bibfnamefont {E.}~\bibnamefont
  {Kiseleva}},\ }\bibfield  {title} {\bibinfo {title} {Kinetic phenomena in
  boundary films of liquids},\ }\href@noop {} {\bibfield  {journal} {\bibinfo
  {journal} {Kolloidn. Zh.}\ }\textbf {\bibinfo {volume} {9}},\ \bibinfo
  {pages} {335} (\bibinfo {year} {1947})}\BibitemShut {NoStop}%
\bibitem [{\citenamefont {Anderson}(1989)}]{Anderson1989}%
  \BibitemOpen
  \bibfield  {author} {\bibinfo {author} {\bibfnamefont {J.~L.}\ \bibnamefont
  {Anderson}},\ }\bibfield  {title} {\bibinfo {title} {Colloid transport by
  interfacial forces},\ }\href@noop {} {\bibfield  {journal} {\bibinfo
  {journal} {Annu. Rev. Fluid Mech.}\ }\textbf {\bibinfo {volume} {21}},\
  \bibinfo {pages} {61} (\bibinfo {year} {1989})}\BibitemShut {NoStop}%
\bibitem [{\citenamefont {Ruckner}\ and\ \citenamefont
  {Kapral}(2007)}]{RuKr07}%
  \BibitemOpen
  \bibfield  {author} {\bibinfo {author} {\bibfnamefont {G.}~\bibnamefont
  {Ruckner}}\ and\ \bibinfo {author} {\bibfnamefont {R.}~\bibnamefont
  {Kapral}},\ }\bibfield  {title} {\bibinfo {title} {Chemically powered
  nanodimers},\ }\href@noop {} {\bibfield  {journal} {\bibinfo  {journal}
  {Phys. Rev. Lett.}\ }\textbf {\bibinfo {volume} {98}},\ \bibinfo {pages}
  {150603:1} (\bibinfo {year} {2007})}\BibitemShut {NoStop}%
\bibitem [{\citenamefont {Golestanian}\ \emph {et~al.}(2007)\citenamefont
  {Golestanian}, \citenamefont {Liverpool},\ and\ \citenamefont
  {Ajdari}}]{Golestanian2007}%
  \BibitemOpen
  \bibfield  {author} {\bibinfo {author} {\bibfnamefont {R.}~\bibnamefont
  {Golestanian}}, \bibinfo {author} {\bibfnamefont {T.~B.}\ \bibnamefont
  {Liverpool}},\ and\ \bibinfo {author} {\bibfnamefont {A.}~\bibnamefont
  {Ajdari}},\ }\bibfield  {title} {\bibinfo {title} {Designing phoretic micro-
  and nano-swimmers},\ }\href@noop {} {\bibfield  {journal} {\bibinfo
  {journal} {New J. Phys.}\ }\textbf {\bibinfo {volume} {9}},\ \bibinfo {pages}
  {126:1} (\bibinfo {year} {2007})}\BibitemShut {NoStop}%
\bibitem [{\citenamefont {Moran}\ and\ \citenamefont
  {Posner}(2017)}]{Moran2017}%
  \BibitemOpen
  \bibfield  {author} {\bibinfo {author} {\bibfnamefont {J.~L.}\ \bibnamefont
  {Moran}}\ and\ \bibinfo {author} {\bibfnamefont {J.~D.}\ \bibnamefont
  {Posner}},\ }\bibfield  {title} {\bibinfo {title} {Phoretic
  self-propulsion},\ }\href@noop {} {\bibfield  {journal} {\bibinfo  {journal}
  {Annu. Rev. Fluid Mech.}\ }\textbf {\bibinfo {volume} {49}},\ \bibinfo
  {pages} {511 } (\bibinfo {year} {2017})}\BibitemShut {NoStop}%
\bibitem [{\citenamefont {W\"urger}(2015)}]{Wurger2015}%
  \BibitemOpen
  \bibfield  {author} {\bibinfo {author} {\bibfnamefont {A.}~\bibnamefont
  {W\"urger}},\ }\bibfield  {title} {\bibinfo {title} {Self-diffusiophoresis of
  {Janus} particles in near-critical mixtures},\ }\href@noop {} {\bibfield
  {journal} {\bibinfo  {journal} {Phys. Rev. Lett.}\ }\textbf {\bibinfo
  {volume} {115}},\ \bibinfo {pages} {188304:1} (\bibinfo {year}
  {2015})}\BibitemShut {NoStop}%
\bibitem [{\citenamefont {Samin}\ and\ \citenamefont {van
  Roij}(2015)}]{samin2015self}%
  \BibitemOpen
  \bibfield  {author} {\bibinfo {author} {\bibfnamefont {S.}~\bibnamefont
  {Samin}}\ and\ \bibinfo {author} {\bibfnamefont {R.}~\bibnamefont {van
  Roij}},\ }\bibfield  {title} {\bibinfo {title} {Self-propulsion mechanism of
  active {Janus} particles in near-critical binary mixtures},\ }\href@noop {}
  {\bibfield  {journal} {\bibinfo  {journal} {Phys. Rev. Lett.}\ }\textbf
  {\bibinfo {volume} {115}},\ \bibinfo {pages} {188305:1} (\bibinfo {year}
  {2015})}\BibitemShut {NoStop}%
\bibitem [{\citenamefont {Brown}\ \emph {et~al.}(2017)\citenamefont {Brown},
  \citenamefont {Poon}, \citenamefont {Holm},\ and\ \citenamefont
  {de~Graaf}}]{Brown2017}%
  \BibitemOpen
  \bibfield  {author} {\bibinfo {author} {\bibfnamefont {A.~T.}\ \bibnamefont
  {Brown}}, \bibinfo {author} {\bibfnamefont {W.~C.~K.}\ \bibnamefont {Poon}},
  \bibinfo {author} {\bibfnamefont {C.}~\bibnamefont {Holm}},\ and\ \bibinfo
  {author} {\bibfnamefont {J.}~\bibnamefont {de~Graaf}},\ }\bibfield  {title}
  {\bibinfo {title} {Ionic screening and dissociation are crucial for
  understanding chemical self-propulsion in polar solvents},\ }\href@noop {}
  {\bibfield  {journal} {\bibinfo  {journal} {Soft Matter}\ }\textbf {\bibinfo
  {volume} {13}},\ \bibinfo {pages} {1200 } (\bibinfo {year}
  {2017})}\BibitemShut {NoStop}%
\bibitem [{\citenamefont {Sharifi-Mood}\ \emph {et~al.}(2013)\citenamefont
  {Sharifi-Mood}, \citenamefont {Koplik},\ and\ \citenamefont
  {Maldarelli}}]{Koplik2013}%
  \BibitemOpen
  \bibfield  {author} {\bibinfo {author} {\bibfnamefont {N.}~\bibnamefont
  {Sharifi-Mood}}, \bibinfo {author} {\bibfnamefont {J.}~\bibnamefont
  {Koplik}},\ and\ \bibinfo {author} {\bibfnamefont {C.}~\bibnamefont
  {Maldarelli}},\ }\bibfield  {title} {\bibinfo {title} {Diffusiophoretic
  self-propulsion of colloids driven by a surface reaction: The sub-micron
  particle regime for exponential and van der {Waals} interactions},\
  }\href@noop {} {\bibfield  {journal} {\bibinfo  {journal} {Phys. Fluids}\
  }\textbf {\bibinfo {volume} {25}},\ \bibinfo {pages} {012001:1} (\bibinfo
  {year} {2013})}\BibitemShut {NoStop}%
\bibitem [{\citenamefont {Sabass}\ and\ \citenamefont
  {Seifert}(2012)}]{Seifert2012}%
  \BibitemOpen
  \bibfield  {author} {\bibinfo {author} {\bibfnamefont {B.}~\bibnamefont
  {Sabass}}\ and\ \bibinfo {author} {\bibfnamefont {U.}~\bibnamefont
  {Seifert}},\ }\bibfield  {title} {\bibinfo {title} {Dynamics and efficiency
  of a self-propelled, diffusiophoretic swimmer},\ }\href@noop {} {\bibfield
  {journal} {\bibinfo  {journal} {J. Chem. Phys.}\ }\textbf {\bibinfo {volume}
  {136}},\ \bibinfo {pages} {064508:1} (\bibinfo {year} {2012})}\BibitemShut
  {NoStop}%
\bibitem [{\citenamefont {Michelin}\ and\ \citenamefont
  {Lauga}(2014)}]{MiLa14}%
  \BibitemOpen
  \bibfield  {author} {\bibinfo {author} {\bibfnamefont {S.}~\bibnamefont
  {Michelin}}\ and\ \bibinfo {author} {\bibfnamefont {E.}~\bibnamefont
  {Lauga}},\ }\bibfield  {title} {\bibinfo {title} {Phoretic self-propulsion at
  finite p{\'e}clet numbers},\ }\href@noop {} {\bibfield  {journal} {\bibinfo
  {journal} {J.~Fluid Mech.}\ }\textbf {\bibinfo {volume} {747}},\ \bibinfo
  {pages} {572} (\bibinfo {year} {2014})}\BibitemShut {NoStop}%
\bibitem [{\citenamefont {Popescu}\ \emph {et~al.}(2016)\citenamefont
  {Popescu}, \citenamefont {Uspal},\ and\ \citenamefont
  {Dietrich}}]{Popescu2016}%
  \BibitemOpen
  \bibfield  {author} {\bibinfo {author} {\bibfnamefont {M.}~\bibnamefont
  {Popescu}}, \bibinfo {author} {\bibfnamefont {W.}~\bibnamefont {Uspal}},\
  and\ \bibinfo {author} {\bibfnamefont {S.}~\bibnamefont {Dietrich}},\
  }\bibfield  {title} {\bibinfo {title} {Self-diffusiophoresis of chemically
  active colloids},\ }\href@noop {} {\bibfield  {journal} {\bibinfo  {journal}
  {Eur. Phys. J. Spec. Top.}\ }\textbf {\bibinfo {volume} {225}},\ \bibinfo
  {pages} {2189 – 2206} (\bibinfo {year} {2016})}\BibitemShut {NoStop}%
\bibitem [{\citenamefont {Domínguez}\ \emph {et~al.}(2020)\citenamefont
  {Domínguez}, \citenamefont {Popescu}, \citenamefont {Rohwer},\ and\
  \citenamefont {Dietrich}}]{DPRD20}%
  \BibitemOpen
  \bibfield  {author} {\bibinfo {author} {\bibfnamefont {A.}~\bibnamefont
  {Domínguez}}, \bibinfo {author} {\bibfnamefont {M.~N.}\ \bibnamefont
  {Popescu}}, \bibinfo {author} {\bibfnamefont {C.~M.}\ \bibnamefont
  {Rohwer}},\ and\ \bibinfo {author} {\bibfnamefont {S.}~\bibnamefont
  {Dietrich}},\ }\bibfield  {title} {\bibinfo {title} {Self-motility of an
  active particle induced by correlations in the surrounding solution},\
  }\href@noop {} {\bibfield  {journal} {\bibinfo  {journal} {Phys. Rev. Lett}\
  }\textbf {\bibinfo {volume} {125}},\ \bibinfo {pages} {268002} (\bibinfo
  {year} {2020})}\BibitemShut {NoStop}%
\bibitem [{\citenamefont {De~Corato}\ \emph {et~al.}(2020)\citenamefont
  {De~Corato}, \citenamefont {Arqu\'e}, \citenamefont {Pati\~no}, \citenamefont
  {Arroyo}, \citenamefont {S\'anchez},\ and\ \citenamefont
  {Pagonabarraga}}]{Corato2020}%
  \BibitemOpen
  \bibfield  {author} {\bibinfo {author} {\bibfnamefont {M.}~\bibnamefont
  {De~Corato}}, \bibinfo {author} {\bibfnamefont {X.}~\bibnamefont {Arqu\'e}},
  \bibinfo {author} {\bibfnamefont {T.}~\bibnamefont {Pati\~no}}, \bibinfo
  {author} {\bibfnamefont {M.}~\bibnamefont {Arroyo}}, \bibinfo {author}
  {\bibfnamefont {S.}~\bibnamefont {S\'anchez}},\ and\ \bibinfo {author}
  {\bibfnamefont {I.}~\bibnamefont {Pagonabarraga}},\ }\bibfield  {title}
  {\bibinfo {title} {Self-propulsion of active colloids via ion release: Theory
  and experiments},\ }\href@noop {} {\bibfield  {journal} {\bibinfo  {journal}
  {Phys. Rev. Lett.}\ }\textbf {\bibinfo {volume} {124}},\ \bibinfo {pages}
  {108001} (\bibinfo {year} {2020})}\BibitemShut {NoStop}%
\bibitem [{\citenamefont {Domínguez}\ and\ \citenamefont
  {Popescu}(2022)}]{DoPo22}%
  \BibitemOpen
  \bibfield  {author} {\bibinfo {author} {\bibfnamefont {A.}~\bibnamefont
  {Domínguez}}\ and\ \bibinfo {author} {\bibfnamefont {M.~N.}\ \bibnamefont
  {Popescu}},\ }\bibfield  {title} {\bibinfo {title} {A fresh view on phoresis
  and self-phoresis},\ }\href@noop {} {\bibfield  {journal} {\bibinfo
  {journal} {Curr. Op. Colloid Interf. Sci.}\ }\textbf {\bibinfo {volume}
  {61}},\ \bibinfo {pages} {101610} (\bibinfo {year} {2022})}\BibitemShut
  {NoStop}%
\bibitem [{\citenamefont {Shreshta}\ and\ \citenamefont {de~la
  Cruz}(2024)}]{Cruz2024}%
  \BibitemOpen
  \bibfield  {author} {\bibinfo {author} {\bibfnamefont {A.}~\bibnamefont
  {Shreshta}}\ and\ \bibinfo {author} {\bibfnamefont {M.}~\bibnamefont {de~la
  Cruz}},\ }\bibfield  {title} {\bibinfo {title} {Enhanced phoretic
  self-propulsion of active colloids through surface charge asymmetry},\
  }\href@noop {} {\bibfield  {journal} {\bibinfo  {journal} {Phys. Rev. E}\
  }\textbf {\bibinfo {volume} {109}},\ \bibinfo {pages} {014613} (\bibinfo
  {year} {2024})}\BibitemShut {NoStop}%
\bibitem [{\citenamefont {Dom{\'i}nguez}\ and\ \citenamefont
  {Popescu}(2024)}]{DoPoion24}%
  \BibitemOpen
  \bibfield  {author} {\bibinfo {author} {\bibfnamefont {A.}~\bibnamefont
  {Dom{\'i}nguez}}\ and\ \bibinfo {author} {\bibfnamefont {M.}~\bibnamefont
  {Popescu}},\ }\bibfield  {title} {\bibinfo {title} {Ionic self-phoresis maps
  onto correlation–-induced self-phoresis},\ }\href@noop {} {\bibfield
  {journal} {\bibinfo  {journal} {arXiv:2404.16435}\ } (\bibinfo {year}
  {2024})}\BibitemShut {NoStop}%
\bibitem [{\citenamefont {Gupta}\ \emph {et~al.}(2019)\citenamefont {Gupta},
  \citenamefont {Rallabandi},\ and\ \citenamefont {Stone}}]{Stone2019b}%
  \BibitemOpen
  \bibfield  {author} {\bibinfo {author} {\bibfnamefont {A.}~\bibnamefont
  {Gupta}}, \bibinfo {author} {\bibfnamefont {B.}~\bibnamefont {Rallabandi}},\
  and\ \bibinfo {author} {\bibfnamefont {H.~A.}\ \bibnamefont {Stone}},\
  }\bibfield  {title} {\bibinfo {title} {Diffusiophoretic and diffusioosmotic
  velocities for mixtures of valence-asymmetric electrolytes},\ }\href@noop {}
  {\bibfield  {journal} {\bibinfo  {journal} {Phys. Rev. Fluids}\ }\textbf
  {\bibinfo {volume} {4}},\ \bibinfo {pages} {043702} (\bibinfo {year}
  {2019})}\BibitemShut {NoStop}%
\bibitem [{\citenamefont {Wilson}\ \emph {et~al.}(2020)\citenamefont {Wilson},
  \citenamefont {Shim}, \citenamefont {Yu}, \citenamefont {Gupta},\ and\
  \citenamefont {Stone}}]{Stone2020}%
  \BibitemOpen
  \bibfield  {author} {\bibinfo {author} {\bibfnamefont {J.~L.}\ \bibnamefont
  {Wilson}}, \bibinfo {author} {\bibfnamefont {S.}~\bibnamefont {Shim}},
  \bibinfo {author} {\bibfnamefont {Y.~E.}\ \bibnamefont {Yu}}, \bibinfo
  {author} {\bibfnamefont {A.}~\bibnamefont {Gupta}},\ and\ \bibinfo {author}
  {\bibfnamefont {H.~A.}\ \bibnamefont {Stone}},\ }\bibfield  {title} {\bibinfo
  {title} {Diffusiophoresis in multivalent electrolytes},\ }\href@noop {}
  {\bibfield  {journal} {\bibinfo  {journal} {Langmuir}\ }\textbf {\bibinfo
  {volume} {36}},\ \bibinfo {pages} {7014 } (\bibinfo {year}
  {2020})}\BibitemShut {NoStop}%
\bibitem [{\citenamefont {Warren}(2020)}]{Warren2020}%
  \BibitemOpen
  \bibfield  {author} {\bibinfo {author} {\bibfnamefont {P.~B.}\ \bibnamefont
  {Warren}},\ }\bibfield  {title} {\bibinfo {title} {Non-faradaic electric
  currents in the {Nernst-Planck} equations and nonlocal diffusiophoresis of
  suspended colloids in crossed salt gradients},\ }\href@noop {} {\bibfield
  {journal} {\bibinfo  {journal} {Phys. Rev. Lett.}\ }\textbf {\bibinfo
  {volume} {124}},\ \bibinfo {pages} {248004} (\bibinfo {year}
  {2020})}\BibitemShut {NoStop}%
\bibitem [{\citenamefont {Williams}\ \emph {et~al.}(2024)\citenamefont
  {Williams}, \citenamefont {Warren}, \citenamefont {Sear},\ and\ \citenamefont
  {Keddie}}]{Warren2024}%
  \BibitemOpen
  \bibfield  {author} {\bibinfo {author} {\bibfnamefont {I.}~\bibnamefont
  {Williams}}, \bibinfo {author} {\bibfnamefont {P.~B.}\ \bibnamefont
  {Warren}}, \bibinfo {author} {\bibfnamefont {R.~P.}\ \bibnamefont {Sear}},\
  and\ \bibinfo {author} {\bibfnamefont {J.~L.}\ \bibnamefont {Keddie}},\
  }\bibfield  {title} {\bibinfo {title} {Colloidal diffusiophoresis in crossed
  electrolyte gradients: {Experimental} demonstration of an
  ``action-at-a-distance'' effect predicted by the nernst-planck equations},\
  }\href@noop {} {\bibfield  {journal} {\bibinfo  {journal} {Phys. Rev.
  Fluids}\ }\textbf {\bibinfo {volume} {9}},\ \bibinfo {pages} {014201}
  (\bibinfo {year} {2024})}\BibitemShut {NoStop}%
\bibitem [{\citenamefont {Happel}\ and\ \citenamefont
  {Brenner}(1965)}]{BrennerBook}%
  \BibitemOpen
  \bibfield  {author} {\bibinfo {author} {\bibfnamefont {J.}~\bibnamefont
  {Happel}}\ and\ \bibinfo {author} {\bibfnamefont {H.}~\bibnamefont
  {Brenner}},\ }\href@noop {} {\emph {\bibinfo {title} {Low Reynolds Number
  Hydrodynamics}}}\ (\bibinfo  {publisher} {Prentice-Hall},\ \bibinfo {address}
  {Englewood Cliffs, NJ},\ \bibinfo {year} {1965})\BibitemShut {NoStop}%
\bibitem [{\citenamefont {Kim}\ and\ \citenamefont {Karrila}(1991)}]{KiKa91}%
  \BibitemOpen
  \bibfield  {author} {\bibinfo {author} {\bibfnamefont {S.}~\bibnamefont
  {Kim}}\ and\ \bibinfo {author} {\bibfnamefont {S.~J.}\ \bibnamefont
  {Karrila}},\ }\href@noop {} {\emph {\bibinfo {title} {Microhydrodynamics:
  Principles and Selected Applications}}}\ (\bibinfo  {publisher}
  {Butterworth--Heinemann},\ \bibinfo {address} {New York},\ \bibinfo {year}
  {1991})\BibitemShut {NoStop}%
\bibitem [{\citenamefont {Lorentz}(1896)}]{Lorentz_original}%
  \BibitemOpen
  \bibfield  {author} {\bibinfo {author} {\bibfnamefont {H.~A.}\ \bibnamefont
  {Lorentz}},\ }\bibfield  {title} {\bibinfo {title} {Eene algemeene stelling
  omtrent de beweging eener vloeistof met wrijving en eenige daaruit afgeleide
  gevolgen},\ }\href@noop {} {\bibfield  {journal} {\bibinfo  {journal}
  {{Zittingsverslag van de Koninklijke Akademie van Wetenschappen te
  Amsterdam}}\ }\textbf {\bibinfo {volume} {5}},\ \bibinfo {pages} {168 }
  (\bibinfo {year} {1896})}\BibitemShut {NoStop}%
\bibitem [{\citenamefont {Kuiken}(1996)}]{Lorentz_transl}%
  \BibitemOpen
  \bibfield  {author} {\bibinfo {author} {\bibfnamefont {H.~K.}\ \bibnamefont
  {Kuiken}},\ }\bibfield  {title} {\bibinfo {title} {{H.A. Lorentz: A general
  theorem on the motion of a fluid with friction and a few results derived from
  it (translated from Dutch by H.K. Kuiken)}},\ }\href@noop {} {\bibfield
  {journal} {\bibinfo  {journal} {J. Eng. Math.}\ }\textbf {\bibinfo {volume}
  {30}},\ \bibinfo {pages} {19 } (\bibinfo {year} {1996})}\BibitemShut
  {NoStop}%
\bibitem [{\citenamefont {Domínguez}\ and\ \citenamefont
  {Popescu}(2024)}]{DP24}%
  \BibitemOpen
  \bibfield  {author} {\bibinfo {author} {\bibfnamefont {A.}~\bibnamefont
  {Domínguez}}\ and\ \bibinfo {author} {\bibfnamefont {M.~N.}\ \bibnamefont
  {Popescu}},\ }\bibfield  {title} {\bibinfo {title} {A representation of the
  velocity of a rigid particle in a forced stokes flow},\ }\href@noop {}
  {\bibfield  {journal} {\bibinfo  {journal} {Phys. Rev. Fluids}\ ,\ \bibinfo
  {pages} {in preparation}} (\bibinfo {year} {2024})}\BibitemShut {NoStop}%
\bibitem [{\citenamefont {de~Groot}\ and\ \citenamefont
  {Mazur}(1984)}]{deMa84}%
  \BibitemOpen
  \bibfield  {author} {\bibinfo {author} {\bibfnamefont {S.~R.}\ \bibnamefont
  {de~Groot}}\ and\ \bibinfo {author} {\bibfnamefont {P.}~\bibnamefont
  {Mazur}},\ }\href@noop {} {\emph {\bibinfo {title} {Non-Equilibrium
  Thermodynamics}}}\ (\bibinfo  {publisher} {Dover},\ \bibinfo {address} {New
  York},\ \bibinfo {year} {1984})\BibitemShut {NoStop}%
\bibitem [{\citenamefont {Anderson}\ \emph {et~al.}(1998)\citenamefont
  {Anderson}, \citenamefont {McFadden},\ and\ \citenamefont
  {Wheeler}}]{Wheeler1998}%
  \BibitemOpen
  \bibfield  {author} {\bibinfo {author} {\bibfnamefont {D.~M.}\ \bibnamefont
  {Anderson}}, \bibinfo {author} {\bibfnamefont {G.~B.}\ \bibnamefont
  {McFadden}},\ and\ \bibinfo {author} {\bibfnamefont {A.~A.}\ \bibnamefont
  {Wheeler}},\ }\bibfield  {title} {\bibinfo {title} {Diffuse-interface methods
  in fluid mechanics},\ }\href@noop {} {\bibfield  {journal} {\bibinfo
  {journal} {Annu. Rev. Fluid Mech.}\ }\textbf {\bibinfo {volume} {30}},\
  \bibinfo {pages} {139 } (\bibinfo {year} {1998})}\BibitemShut {NoStop}%
\bibitem [{\citenamefont {Teubner}(1982)}]{Teubner1982}%
  \BibitemOpen
  \bibfield  {author} {\bibinfo {author} {\bibfnamefont {M.}~\bibnamefont
  {Teubner}},\ }\bibfield  {title} {\bibinfo {title} {Motion of charged
  colloidal particles},\ }\href@noop {} {\bibfield  {journal} {\bibinfo
  {journal} {J. Chem. Phys.}\ }\textbf {\bibinfo {volume} {76}},\ \bibinfo
  {pages} {5564} (\bibinfo {year} {1982})}\BibitemShut {NoStop}%
\bibitem [{\citenamefont {Kim}\ \emph {et~al.}(2017)\citenamefont {Kim},
  \citenamefont {Choudhury}, \citenamefont {Hyeon-Ho},\ and\ \citenamefont
  {Fischer}}]{Fischer2017}%
  \BibitemOpen
  \bibfield  {author} {\bibinfo {author} {\bibfnamefont {J.~T.}\ \bibnamefont
  {Kim}}, \bibinfo {author} {\bibfnamefont {U.}~\bibnamefont {Choudhury}},
  \bibinfo {author} {\bibfnamefont {J.}~\bibnamefont {Hyeon-Ho}},\ and\
  \bibinfo {author} {\bibfnamefont {P.}~\bibnamefont {Fischer}},\ }\bibfield
  {title} {\bibinfo {title} {Nanodiamonds that swim},\ }\href
  {https://doi.org/10.1002/adma.201701024} {\bibfield  {journal} {\bibinfo
  {journal} {Adv. Mater.}\ }\textbf {\bibinfo {volume} {29}},\ \bibinfo {pages}
  {1701024} (\bibinfo {year} {2017})}\BibitemShut {NoStop}%
\bibitem [{\citenamefont {Hortelao}\ \emph {et~al.}(2021)\citenamefont
  {Hortelao}, \citenamefont {Sim{\'o}}, \citenamefont {Guix}, \citenamefont
  {Guallar-Garrido}, \citenamefont {Juli{\'a}n}, \citenamefont {Vilela},
  \citenamefont {Rejc}, \citenamefont {Ramos-Cabrer}, \citenamefont {Cossío},
  \citenamefont {G{\'o}mez-Vallejo}, \citenamefont {Pati{\~n}o}, \citenamefont
  {Llop},\ and\ \citenamefont {S{\'a}nchez}}]{Sanchez2021}%
  \BibitemOpen
  \bibfield  {author} {\bibinfo {author} {\bibfnamefont {A.~C.}\ \bibnamefont
  {Hortelao}}, \bibinfo {author} {\bibfnamefont {C.}~\bibnamefont {Sim{\'o}}},
  \bibinfo {author} {\bibfnamefont {M.}~\bibnamefont {Guix}}, \bibinfo {author}
  {\bibfnamefont {S.}~\bibnamefont {Guallar-Garrido}}, \bibinfo {author}
  {\bibfnamefont {E.}~\bibnamefont {Juli{\'a}n}}, \bibinfo {author}
  {\bibfnamefont {D.}~\bibnamefont {Vilela}}, \bibinfo {author} {\bibfnamefont
  {L.}~\bibnamefont {Rejc}}, \bibinfo {author} {\bibfnamefont {P.}~\bibnamefont
  {Ramos-Cabrer}}, \bibinfo {author} {\bibfnamefont {U.}~\bibnamefont
  {Cossío}}, \bibinfo {author} {\bibfnamefont {V.}~\bibnamefont
  {G{\'o}mez-Vallejo}}, \bibinfo {author} {\bibfnamefont {T.}~\bibnamefont
  {Pati{\~n}o}}, \bibinfo {author} {\bibfnamefont {J.}~\bibnamefont {Llop}},\
  and\ \bibinfo {author} {\bibfnamefont {S.}~\bibnamefont {S{\'a}nchez}},\
  }\bibfield  {title} {\bibinfo {title} {Swarming behavior and in vivo
  monitoring of enzymatic nanomotors within the bladder},\ }\href
  {https://doi.org/10.1126/scirobotics.abd2823} {\bibfield  {journal} {\bibinfo
   {journal} {Sci. Robotics}\ }\textbf {\bibinfo {volume} {6}},\ \bibinfo
  {pages} {eabd2823} (\bibinfo {year} {2021})},\ \Eprint
  {https://arxiv.org/abs/https://www.science.org/doi/pdf/10.1126/scirobotics.abd2823}
  {https://www.science.org/doi/pdf/10.1126/scirobotics.abd2823} \BibitemShut
  {NoStop}%
\end{thebibliography}
%

\appendix

\section{Spatial variation of $\delta\mu$}
\label{app:mu}

The Laplace equation~(\ref{eq:muLaplace}) can be obtained from a
variational principle as an extremum of the functional
\begin{equation}
  \label{eq:3}
  \mathcal{S}[\mu] = \int\limits_{r>R} d^3\br\; \frac{1}{2} |\nabla
  \mu|^2
  = \int\limits_{r>R} d^3\br\; \frac{1}{2} \left[
    \left(\frac{\partial\mu}{\partial r}\right)^2
    + |\nabla_\parallel \mu|^2
  \right],
\end{equation}
so that, for non-trivial solutions, one may expect
$|\partial_r\mu| \sim |\nabla_\parallel\mu|$ in order of
magnitude. Alternatively, one can estimate these magnitudes thorough
the surface average of their squares evaluated with the explicit
solution~(\ref{eq:mulm}) on the surface ($z=0$). It follows directly
from the boundary condition~(\ref{eq:neumannMu}) that 
\begin{equation}
  \left\langle |\partial_r \mu|^2\right\rangle
  = \left(\frac{\mathcal{A}}{n_0 \Gamma} \right)^2
  \left\langle |\mathbb{A}|^2 \right\rangle
  = \left(\frac{\mathcal{A}}{n_0 \Gamma} \right)^2
  \sum_{\ell m} \frac{|a_{\ell m}|^2}{4\pi} .
\end{equation}
For the lateral variation, one has
\begin{equation}
  \label{eq:nablaparMu}
  \left\langle |\nabla_\parallel \mu|^2\right\rangle
  = \left(\frac{\mathcal{A} R}{n_0 \Gamma} \right)^2
  \sum_{\ell m} \sum_{\ell' m'}
  \nu_{\ell m}(0) \nu_{\ell' m'}(0)
  \left\langle
    \left(\nabla_\parallel Y_{\ell m}^*\right)
    \cdot\left(\nabla_\parallel Y_{\ell' m'}\right)
  \right\rangle .
\end{equation}
One can exploit the properties of the spherical harmonics
(orthonormality and eigenfunctions of $\nabla_\parallel^2$):
\begin{eqnarray}
  \left\langle
    \left(\nabla_\parallel Y_{\ell m}^*\right)
    \cdot
    \left(\nabla_\parallel Y_{\ell' m'}\right)
  \right\rangle
  & =
  & \left\langle \nabla_\parallel\cdot \left(
      Y_{\ell m}^*
      \nabla_\parallel Y_{\ell' m'}
    \right)
  \right\rangle
  - \left\langle Y_{\ell m}^*
      \nabla_\parallel^2 Y_{\ell' m'}
    \right\rangle
    \nonumber
  \\
  & =
  & \frac{\ell' (\ell'+1)}{R^2}
    \left\langle
    Y_{\ell m}^* \, Y_{\ell' m'}
    \right\rangle
    = \frac{\ell' (\ell'+1)}{4\pi R^2} \delta_{\ell \ell'}
    \delta_{m m'} ,
\end{eqnarray}
where in the first line the first average vanishes by virtue of Gauss theorem on curved manifolds. Therefore, \eq{eq:nablaparMu} becomes
\begin{equation}
  \left\langle |\nabla_\parallel \mu|^2\right\rangle
  = \left(\frac{\mathcal{A}}{n_0 \Gamma} \right)^2
  \sum_{\ell m} \frac{\ell}{\ell+1} \frac{|a_{\ell m}|^2}{4\pi} ,
\end{equation}
after using \eq{eq:mucoeff}, and one can estimate the ratio
\begin{equation}
  \frac{\left\langle |\nabla_\parallel
      \mu|^2\right\rangle}
  {\left\langle |\partial_r \mu|^2\right\rangle}
  = 1 - \frac{\sum_{\ell m} (\ell+1)^{-1} |a_{\ell m}|^2}{\sum_{\ell m} |a_{\ell m}|^2} ,
\end{equation}
which will be typically of order unity, particularly when the sum is
dominated by small scales of lateral variation ($\ell\gg 1$).

\section{The ideal gas}
\label{app:idealgas}

For an ideal gas, the following relationship holds according to
Eqs.~(\ref{eq:mun}, \ref{eq:hideal}):
\begin{equation}
  \label{eq:Zideal}
  n = \mathrm{exp} \left(\frac{\mu-\mathbb{W}}{kT} \right),
  \qquad
  h''(n)=\frac{kT}{n} .
\end{equation}
Therefore, one can write the integrand in \eq{eq:vs_lubr1} as
\begin{equation}
  \frac{1}{h''(n)}
  \frac{\partial \mathbb{W}}{\partial z}
  \nabla_\parallel \mu
  = k T  \frac{\partial}{\partial z} \left( 1 - \mathrm{e}^{-\mathbb{W}/kT}
  \right)\, \nabla_\parallel \mathrm{e}^{\mu/kT} .
\end{equation}
One can now argue as in Sec.~\ref{sec:lubr}: if the integral is
effectively cut off at a distance of the order of the range $\sigma$
of $\mathcal{W}$, and one can argue that $\mu$ varies over a much
larger length scale, one can approximate
\begin{equation}
  \bv_s(\br_p) \approx \mathcal{L}_\mathrm{id}(\br_p)
  \; \nabla_\parallel \mathrm{e}^{\mu (\br_p)/kT} ,
\end{equation}
with
\begin{equation}
  \label{eq:Lideal1}
  \mathcal{L}_\mathrm{id}(\br_p)
  :=
  - \frac{kT}{2\eta}\int\limits_0^\infty dz\; z^2\,
  \frac{\partial}{\partial z} \left( 1 - \mathrm{e}^{-\mathbb{W}/kT}
  \right) .
\end{equation}
Outside of the thin layer, where $\mathbb{W}\approx 0$, one can define
the ``outer density'' as $n^\mathrm{(o)}=\mathrm{e}^{\mu (\br_p)/kT}$
by \eq{eq:Zideal}, so that one arrives at \eq{eq:vs_lubr_quasihom}
with the phoretic coefficient given by \eq{eq:Lclassic} after
integrating by parts in \eq{eq:Lideal1}.

\section{Expansion in spherical harmonics}
\label{app:Yexp}

When the definitions~(\ref{eq:V_Om_curlf}) are used to compute the
surface averages~(\ref{eq:slips}), one gets
\begin{subequations}
  \label{eq:VOm_exact}
  \begin{equation}
    \bV = \frac{2}{3\eta R} \int\limits_{0}^{\infty} dz\; (R+z)^2\,
    A(R+z) \, \langle \be_r \times \left( \nabla \times
      \bff
    \right) \rangle,
  \end{equation}
  \begin{equation}
    \bOmega = \frac{2}{\eta R^3}
    \int\limits_{0}^{\infty} dz\; (R+z)^2\,
    \Phi(R+z) \, \langle \nabla \times \bff
    \rangle ,
  \end{equation}
\end{subequations}
since the dependence on the position $\be_r$ on the particle's surface
enters only through $\nabla\times\bff$. It will turn out convenient to redefine the tangential gradient $\nabla_\parallel$, by making explicit the dependence on the radial
coordinate $r$, as 
\begin{equation}
  \label{eq:nablapar}
  \tilde{\nabla}_\parallel := r \nabla_\parallel = \be_\theta
  \partial_\theta + \frac{\be_\varphi}{\sin\theta} \partial_\varphi
\end{equation}
in spherical coordinates. With this redefinition, one can approximate
\eq{eq:curl_f_id} in the quasi-homogeneous approximation as
\begin{eqnarray}
  \label{eq:curl_f_quasihom}
  \nabla \times \bff
  & \approx
  & \frac{1}{h''(n_0)} \nabla \mathbb{W} \times \nabla \mu
    \nonumber \\
  & =
  & \frac{1}{h''(n_0) r^2} \left[
    r \frac{\partial \mathbb{W}}{\partial r} \be_r \times
    \tilde{\nabla}_\parallel \mu
    - r \frac{\partial \mu}{\partial r} \be_r \times
    \tilde{\nabla}_\parallel \mathbb{W}
    + \tilde{\nabla}_\parallel \mathbb{W} \times
    \tilde{\nabla}_\parallel\mu
  \right] ,
\end{eqnarray}
so that one also has
\begin{equation}
  \be_r\times\left(\nabla\times\bff\right)
  \approx
  \frac{1}{h''(n_0) r} \left[
    \frac{\partial \mu}{\partial r}\,
    \tilde{\nabla}_\parallel \mathbb{W}
    - \frac{\partial \mathbb{W}}{\partial r}\,
    \tilde{\nabla}_\parallel \mu
  \right] .
\end{equation}
These fields can be evaluated by inserting the
expansions~(\ref{eq:mulm}, \ref{eq:wlm}), and when the result is used
in \eqs{eq:VOm_exact}, one gets the following expansions:
\begin{subequations}
  \label{eq:VOm_exact2}
  \begin{eqnarray}
    \label{eq:Vexact2}
  \bV = \frac{2 \mathcal{A} \mathcal{W}}{3\eta \Gamma n_0 h''(n_0)}
  & \displaystyle \sum_{\ell m} \sum_{\ell' m'}
  & \left\{
    \left\langle
    Y_{\ell' m'} \tilde{\nabla}_\parallel Y_{\ell m}
    \right\rangle
    \int\limits_{0}^{\infty} dz\; (R+z)\,
    A(R+z)
    \frac{d \mulm_{\ell' m'}}{d z} \, w_{\ell m}(z)
    \right.
  \\
  &
  & - \left.
    \left\langle
    Y_{\ell m} \tilde{\nabla}_\parallel Y_{\ell' m'}
  \right\rangle
  \int\limits_{0}^{\infty} dz\; (R+z)\,
    A(R+z)
    \frac{d w_{\ell m}}{d z} \, \mulm_{\ell' m'}(z)
    \right\} ,
    \nonumber
\end{eqnarray}
\begin{eqnarray}
  \label{eq:Omexact2}
  \bOmega
  = \frac{2 \mathcal{A} \mathcal{W}}{\eta \Gamma n_0 h''(n_0) R^2}
  & \displaystyle \sum_{\ell m} \sum_{\ell' m'}
  & \left\{
    \left\langle
    Y_{\ell m} \; \be_r\times \tilde{\nabla}_\parallel Y_{\ell' m'}
    \right\rangle
    \int\limits_{0}^{\infty} dz\; (R+z) \,
    \Phi(R+z)
    \frac{d w_{\ell m}}{d z} \, \mulm_{\ell' m'}(z)
    \right.
    \\
  &
  & 
    - \left\langle
    Y_{\ell' m'} \; \be_r\times \tilde{\nabla}_\parallel Y_{\ell m}
    \right\rangle
    \int\limits_{0}^{\infty} dz\; (R+z) \,
    \Phi(R+z)
    \frac{d \mulm_{\ell' m'}}{d z} \, w_{\ell m}(z)
    \nonumber
  \\
  &
  & \left.
    + \left\langle
    \left(\tilde{\nabla}_\parallel Y_{\ell m}\right)
    \times
    \left(\tilde{\nabla}_\parallel Y_{\ell' m'}\right)
    \right\rangle
    \int\limits_{0}^{\infty} dz\; 
    \Phi(R+z) \;
    w_{\ell m}(z) \, \mulm_{\ell' m'}(z)
    \right\} .
    \nonumber
\end{eqnarray}
\end{subequations}
Consider first \eq{eq:Vexact2}: using the
definitions~(\ref{eq:mucoeff}, \ref{eq:V0}, \ref{eq:Gpar}) and one can
identify from the first summand the coefficient
\begin{equation}
  \label{eq:c_lm2}
  c_{\ell m; \ell'} := (\ell +1) \int\limits_0^\infty d\zeta \;
  \frac{A(R(1+\zeta))}{R} \, (1+\zeta)^{-\ell'-1} \, w_{\ell m}(R\zeta) ,
\end{equation}
appearing in \eq{eq:V_lm}, in terms of the integration variable
$\zeta := z/R$. In the second summand one can integrate by parts
(notice that the hydrodynamic kernel $A(R+z)$ vanishes at $z=0$ and
that, by assumption, the coefficients $w_{\ell m}(z)$ vanish at
$z\to\infty$ sufficiently fast), so that only the derivatives of
$\mulm_{\ell' m'}$ appear,
\begin{equation}
  \frac{d \mulm_{\ell'm'}}{dz} 
  = - \frac{a_{\ell' m'}}{R} \left( 1 +\frac{z}{R} \right)^{-(\ell'+2)} .
\end{equation}
In this manner one identifies the other coefficient,
\begin{equation}
  \label{eq:b_lm2}
  b_{\ell m; \ell'} := - \int\limits_0^\infty d\zeta \;
  \frac{d}{d\zeta}\left[ (1+\zeta) \frac{A(R(1+\zeta))}{R} \right] \;
  (1+\zeta)^{-\ell'-1} \, w_{\ell m}(R\zeta) .
\end{equation}
By inserting the form~(\ref{eq:Aspher}) of the hydrodynamic kernel in
these expressions, one eventually arrives at \eqs{eq:g_lm} with the
functions given by \eqs{eq:Khydro}.

Addressing now \eq{eq:Omexact2}, one notices the following identities
(see App.~\ref{app:Sidentities}):
\begin{subequations}
  \label{eq:YYaverages}
  \begin{equation}
    \label{eq:YY1}
  \left\langle
    Y_{\ell' m'} \; \be_r\times \tilde{\nabla}_\parallel Y_{\ell m}
  \right\rangle
  = - \left\langle
    Y_{\ell m} \; \be_r\times \tilde{\nabla}_\parallel Y_{\ell' m'}
  \right\rangle ,
\end{equation}
\begin{equation}
  \label{eq:YY2}
  \left\langle
    \left(\tilde{\nabla}_\parallel Y_{\ell m}\right)
    \times
    \left(\tilde{\nabla}_\parallel Y_{\ell' m'}\right)
  \right\rangle
  = \frac{1}{R} \left\langle
    Y_{\ell m} \; \be_r\times \tilde{\nabla}_\parallel Y_{\ell' m'}
  \right\rangle .
\end{equation}
\end{subequations}
Using them together with the definitions~(\ref{eq:mucoeff},
\ref{eq:V0}, \ref{eq:Gtau}) and integration by parts as before, one
obtains
\begin{equation}
  \bOmega = \frac{3 V_0}{2 R}
  \sum_{\ell m} \sum_{\ell' m'}
  a_{\ell' m'} d_{\ell m; \ell'}
  \bG^\tau_{\ell m; \ell' m'},
\end{equation}
with coefficients
\begin{equation}
  d_{\ell m; \ell'} := \int\limits_0^\infty d\zeta \;
  \left\{
    \frac{d}{d\zeta}\left[ (1+\zeta) \frac{\Phi(R(1+\zeta))}{R^2}
      - \frac{\Phi(R(1+\zeta))}{R^2}
    \right]
  \right\} 
  (1+\zeta)^{-\ell'-1} \, w_{\ell m}(R\zeta) .
\end{equation}
This expression can be evaluated by inserting the
form~(\ref{eq:Qspher}) of the hydrodynamic kernel. Against
appearances, it is not independent of the
coefficients~(\ref{eq:c_lm2}, \ref{eq:b_lm2}) and it can be expressed
as a linear combination of them:
\begin{equation}
  \label{eq:dlm}
  d_{\ell m; \ell'} = \int\limits_0^\infty d\zeta \; \left[
    1 - \frac{1}{(1+\zeta)^3} \right]
  (1+\zeta)^{1-\ell'} w_{\ell m}(\zeta)
  =b_{\ell m; \ell'-1} + \frac{c_{\ell m; \ell'-1}}{\ell+1},
\end{equation}
and so is \eq{eq:Om_lm} recovered.

\section{Some integral identities}
\label{app:Sidentities}

Here we derive some useful identities for averages over the surface of
a sphere. In this appendix, $f(\be_r)$, $g(\be_r)$ will denote any two
smooth functions defined on the sphere $\mathbb{S}_2$. The first
identity follows from Stokes theorem for a scalar field:
\begin{equation}
  \int\limits_{\mathbb{S}_2} dS\; \be_r\times\nabla_\parallel f
  = \int\limits_{\mathbb{S}_2} d\bS\times\nabla f
  = \oint\limits_{\partial\mathbb{S}_2} d\boldsymbol{\ell} \; f
  = 0 ,
\end{equation}
taking into account that a sphere is a closed surface, i.e., it has no
boundary: $\partial\mathbb{S}_2 = \emptyset$. Therefore, one has
\begin{equation}
  \langle g\, \be_r\times\nabla_\parallel f \rangle
  = \langle \be_r\times\nabla_\parallel (g f) \rangle
  - \langle f \,\be_r\times\nabla_\parallel g \rangle
  = - \langle f \, \be_r\times\nabla_\parallel g \rangle.
\end{equation}
This identity demonstrates \eq{eq:YY1}. Incidentally, this
implies that $\bG^\tau_{\ell m; \ell m}=0$.

In order to proof \eq{eq:YY2} one uses the explicit
representation~(\ref{eq:nablapar}) and integrates by parts on the surface, accounting that the fields are
assumed to be well defined on the whole sphere, and thus must be
periodic, in order to drop boundary terms:
\begin{eqnarray}
  \langle (\nabla_\parallel g) \times (\nabla_\parallel f) \rangle
  & =
  & \frac{1}{r^2} \left\langle
  \frac{\be_r}{\sin\theta} \left[
    (\partial_\theta g) (\partial_\varphi f)
    - (\partial_\varphi g) (\partial_\theta f)
  \right]
    \right\rangle
    \nonumber
  \\
  & =
  & \frac{1}{r^2} \left\langle
    \be_\varphi \, g \, (\partial_\theta f)
    - \frac{\be_\theta}{\sin\theta} g \, (\partial_\varphi f)
    \right\rangle
    = \frac{1}{r} \left\langle
    g\; \be_r \times\nabla_\parallel f
    \right\rangle .
\end{eqnarray}
Finally, the following identity can be proofed also by explicit
evaluation:
\begin{equation}
  \langle \nabla_\parallel f \rangle
  = \frac{2}{r} \langle \be_r \, f \rangle .
\end{equation}
As a consequence, it holds that
\begin{equation}
  \left\langle
    Y_{\ell m} \nabla_\parallel Y_{\ell' m'}
  \right\rangle
  = \frac{2}{R} \left\langle
    \be_r Y_{\ell m} Y_{\ell' m'}
  \right\rangle
  - \left\langle
    Y_{\ell' m'} \nabla_\parallel Y_{\ell m}
  \right\rangle .  
\end{equation}
This yields the following equality:
\begin{equation}
  \label{eq:Gperp}
  - (\ell'+1)\bG^\parallel_{\ell m; \ell' m'}
  = 2 \bG^\perp_{\ell m; \ell' m'} 
  + (\ell +1) \bG^\parallel_{\ell' m'; \ell m} ,
\end{equation}
in terms of the vectors
\begin{equation}  
  \bG^\perp_{\ell m; \ell' m'} := 4\pi \left\langle
    \be_r Y_{\ell m} Y_{\ell' m'} \right\rangle
\end{equation}
introduced in \rcite{DPRD20}.

\end{document}